\documentclass[12pt]{article}
\usepackage{amsfonts,graphicx,amsmath,amsthm,amssymb,epsfig,color,float}
\allowdisplaybreaks
\begin{document}

\title{\bf Complexity and Isotropization based Extended Models in the context
of Electromagnetic Field: An Implication of Minimal Gravitational
Decoupling}
\author{Tayyab Naseer$^{1,2}$ \thanks{tayyab.naseer@math.uol.edu.pk;tayyab.naseer@khazar.org;tayyabnaseer48@yahoo.com}\\
$^1$Department of Mathematics and Statistics, The University of Lahore,\\
1-KM Defence Road Lahore-54000, Pakistan.\\
$^2$Research Center of Astrophysics and Cosmology, Khazar University, \\
Baku, AZ1096, 41 Mehseti Street, Azerbaijan.}

\date{}
\maketitle

\begin{abstract}
This paper formulates three different analytical solutions to the
gravitational field equations in the framework of Rastall theory by
taking into account the gravitational decoupling approach. For this,
the anisotropic spherical interior fluid distribution is assumed as
a seed source characterized by the corresponding Lagrangian. The
field equations are then modified by introducing an additional
source which is gravitationally coupled with the former fluid setup.
Since this approach makes the Rastall equations more complex, the
MGD scheme is used to tackle this, dividing these equations into two
systems. Some particular ansatz are taken into account to solve the
first system, describing initial anisotropic fluid. These metric
potentials contain multiple constants which are determined with the
help of boundary conditions. On the other hand, the solution for the
second set is calculated through different well-known constraints.
Afterwards, the estimated data of a pulsar $4U 1820-30$ is
considered so that the feasibility of the developed models can be
checked graphically. It is concluded that all resulting models show
physically acceptable behavior under certain choices of Rastall and
decoupling parameters.
\end{abstract}

\section{Introduction}

In 1972, Peter Rastall proposed the notion that the energy-momentum
tensor (EMT) having null divergence in flat spacetime may not
necessarily vanish in curved spacetime configurations \cite{1}.
Since this is the fundamental assumption on which the proposed
modification of general theory of relativity (GR) is based, it named
the Rastall gravity theory. In this theory, distinct from the
Einstein's GR, the inclusion of the Ricci scalar $R$ is achieved
through the introduction of the Rastall parameter. This alteration
fundamentally reshapes the interaction between matter fields and
gravitational forces, leading to a paradigm shift in the
understanding of these phenomena. Within this theoretical framework,
emphasis is placed on non-minimal couplings as the primary mechanism
governing the dynamics of gravity and matter interactions. When
exploring this gravity theory, scientists determined that it stands
on par with other modifications of GR obtained by adjusting the
Einstein-Hilbert action \cite{2}-\cite{7c}.

As per the observational status of this theory is concerned, Batista
et al. \cite{aa} explored the Rastall gravity within the context of
FLRW metric and studied the evolution of small perturbations. They
concluded that the two developed models dramatically stable under
certain parametric values. Fabris et al. \cite{ab} have explored the
implications of this gravity within a cosmological framework,
utilizing data from Type Ia supernovae to investigate the viability
of the model. Their findings indicated that while this extended
theory can accommodate certain cosmological observations, it does
not strictly constrain the parameters involved, preferring values
that align closely with those predicted by GR. Darabi and his
colleagues \cite{ac} argued that the Rastall theory has a
non-minimal coupling that could produce interesting results. An
intriguing aspect of this theory is its ability to uphold any
solution derived from the Einstein's equations. Notably, when
examining black hole configurations, it is explored that both GR and
Rastall gravity yield identical outcomes in scenarios involving
vacuum spacetime. Some interesting works can be found in
\cite{ad}-\cite{aj}.

The celestial structures are typically defined by a set of up to
forth order partial differential equations incorporating geometric
and matter terms, known as the gravitational field equations. The
solutions (extracted either analytically or numerically) of these
complex equations have garnered significant attention among
astrophysicists in recent times. These solutions enable the
evaluation of whether the stellar object being studied possesses
physical relevance or not. To address the needs of astronomers,
diverse methods have been developed and applied to generate
solutions that might be of interest to them \cite{ak}-\cite{aq}. A
notable strategy has been proposed by a research team led by Jorge
Ovalle, known as gravitational decoupling. This approach is
particularly prominent because it can be employed even when multiple
physical factors are involved in the interior geometry. A detailed
outline of this method shall be discussed later in the section where
it is more relevant. The techniques used to analyze this problem can
be broadly categorized into two main approaches. The first method,
known as the minimal geometric deformation (MGD), has been pioneered
by Ovalle \cite{29} which was extended in \cite{30}, who
successfully applied it to obtain exact solutions. This approach was
subsequently employed by Casadio et al. \cite{31}, which led to the
development of the Schwarzschild spacetime.

Ovalle et al. \cite{33} incorporated a new fluid source into the
initially considered matter composition through a Lagrangian
density. This modification led to the generation of novel and
promising findings. The study was subsequently extended to include
the effects of the Maxwell field. In this case, two Lagrangian
densities were considered: one corresponding to the initial source
and another representing the electromagnetic field. Subsequently, a
third Lagrangian density analogous to the additional source was
introduced, further expanding the scope of the investigation and its
potential implications \cite{34}. In an effort to explore this
strategy, alternative theories were investigated, yielding multiple
viable outcomes \cite{35}-\cite{35b}. Gabbanelli et al. \cite{36}
pursued a similar approach by using Durgapal-Fuloria metric as an
original perfect fluid source and developed a physically existing
extension. Following the same, researchers have expanded the domain
of Tolman VII and Heintzmann isotropic models \cite{36a,37a}. Sharif
and Naseer \cite{37b,37f} have significantly broadened the scope of
Ovalle's work by incorporating both charged and uncharged scenarios
in Einstein's as well as modified frameworks, and reported stable
outcomes. Some other interesting works in this regard are
\cite{37x}-\cite{37ya}.

The scientific community has also devoted significant attention to
investigating the factors that contribute to the intricate nature of
celestial objects. Numerous definitions have been put forth over
time to encapsulate this concept, yet they have been deemed
insufficient in certain scenarios. In this regard, Herrera
\cite{37g}, a prominent astrophysicist, recently pioneered the
widely accepted definition of complexity. Following his initial
work, Herrera and his research team later expanded the definition to
encompass situations where heat dissipation plays a crucial role as
a dominant factor in the interior spacetime \cite{37h}. The
underlying motivation for this definition is rooted in orthogonally
decomposing the curvature tensor, yielded multiple scalars that are
interconnected with various physical parameters. Notably, the
primary contributors to the complexity of the system (i.e., density
inhomogeneity and pressure anisotropy) are encapsulated within a
single scalar, which is thus designated as the complexity factor.
Some other factors are also appeared while decomposing the forth
rank Riemann tensor, however, they possess incomplete information to
be called the complexity factor for the considered fluid
configuration.

Recent studies have explored the definition of complexity within the
framework of modified theories \cite{37i}-\cite{37ia}. All of their
results indicated that the factor $Y_{TF}$ remains the complexity
factor, no matters what kind of interior geometry is being studied,
either, static/non-static, uncharged/charged, etc. A notable aspect
of this study is the reliance on specific conditions to solve the
field equations. One such constraint, the vanishing complexity, has
garnered significant attention from researchers in recent years.
This constraint has been effectively employed in the context of
gravitational decoupling to investigate compact models
\cite{37j}-\cite{37jb}. Casadio with his colleagues \cite{37k}
proposed a couple of theoretical constraints, with one model
formulated under disappearing complexity condition and the other
focusing on the isotropization mechanism. Maurya et al. \cite{37l}
subsequently delved into these models through the Karmarkar
condition, investigating the influence of various parameters on
internal complexity and anisotropy in principal pressure. Extending
their analysis to incorporate the impact of Brans-Dicke theory, they
derived physically meaningful solutions \cite{37m}.

The solutions to the gravitational equations play a pivotal role in
the study of compact objects in GR as well as modified theories of
gravity. These solutions provide a framework for modeling the
structure and stability of anisotropic neutron stars, which are
believed to exhibit complex internal dynamics due to factors such as
superfluidity and strong magnetic fields. Recent literature has
highlighted the relevance of the Krori-Barua metric in various
gravitational contexts, demonstrating its applicability in
constructing stable celestial configurations \cite{38a}-\cite{38d}.
Additionally, the Tolman IV solution has been utilized to analyze
anisotropic relativistic spheres in modified gravity theories,
revealing that these models can satisfy essential physical
requirements and avoid singularities \cite{38e}-\cite{38h}. By
leveraging these solutions, researchers can better understand the
intricate balance between gravitational forces and internal
pressures in compact stars, providing valuable constraints on their
mass and radius based on observational data. Consequently, the
utilization of both these spacetimes into astrophysical models not
only enhances our understanding of stellar structures but also
contributes to the ongoing discourse on the viability of Rastall
theory, thereby paving the way for future research in the field.

This article explores the effects of specific parameters on the
newly derived charged models within the framework of Rastall
gravity. To address the corresponding field equations, various
constraints are applied, drawing from the idea of isotropization and
complexity. The structure of this paper is outlined below. The
origin of Rastall theory and the inclusion of an additional source
term are explained in the following section. Section \textbf{3}
outlines the MGD approach in detail and also explains its
implementation. In sections \textbf{4} and \textbf{5}, I derive
three innovative solutions using analytical methods. These solutions
are also visualized and discussed through graphical representations.
Finally, section \textbf{6} provides a summary of the results
obtained within the present scenario. \textcolor{blue} {As a
statement of research hypothesis, the Rastall theory, when applied
to spherically symmetric spacetime with anisotropic pressures,
yields physically viable models that can describe the structure and
behavior of neutron stars, providing insights that extend beyond
those offered by GR.}

\section{Fundamentals of Rastall Theory and Static Spherical Spacetime}

General relativity is based on the fundamental concept of the
conservation of EMT within the spacetime geometry under discussion.
When this result does not hold within a curved spacetime scenario,
i.e., the EMT becomes non-conserved, Rastall theory come into being
\cite{1}. The gravitational equations under respective modification
become
\begin{equation}\label{g1}
G_{\eta\mu}\equiv R_{\eta\mu}-\frac{1}{2}Rg_{\eta\mu}=\kappa
\big(T_{\eta\mu}-\xi Rg_{\eta\mu}\big),
\end{equation}
where
\begin{itemize}
\item $R_{\eta\mu}$ and $G_{\eta\mu}$ denote the Ricci tensor and Einstein
tensor, respectively,
\item $T_{\eta\mu}$ symbolizes the fluid distribution and $\kappa$ being the coupling constant,
\item $\xi$ refers to the Rastall parameter under the control of which this theory differs
from the standard GR.
\end{itemize}
It is exciting to know that there is a consistency between the field
equations \eqref{g1} and the following non-conservation phenomenon
defined as
\begin{equation}\label{g2}
\nabla^\eta T_{\eta\mu}=\xi g_{\eta\mu}\nabla^\eta R,
\end{equation}
which obviously results in the conservation equation for $\xi=0$. It
must also be stated that the so-called non-minimal interaction
between matter and spacetime can be observed on the evolution of
celestial objects under the above non-zero divergence. If we define
the factor
\begin{equation}\label{g3}
\hat{T}_{\eta\mu}=T_{\eta\mu}-\xi Rg_{\eta\mu},
\end{equation}
this helps in redefining the equations of motion \eqref{g1} in the
following
\begin{eqnarray}\label{g4}
G_{\eta\mu}=\kappa \hat{T}_{\eta\mu},
\end{eqnarray}
and, hence, they are compatible with the mathematical result
expressed by $\nabla^\eta \hat{T}_{\eta\mu}=0$. We also observe that
the same result can be obtained for extended theories which are
achieved through the modification of the action function.

To reframe the field equations \eqref{g1}, we begin by taking their
trace. This yields
\begin{eqnarray}\label{g4a}
R=\frac{\kappa T}{4\xi\kappa-1},
\end{eqnarray}
which after substituting this into Eq.\eqref{g3} gives the effective
matter part as
\begin{equation}\label{g4b}
\bar{T}_{\eta\mu}=T_{\eta\mu}-\frac{\beta T}{4\beta-1} g_{\eta\mu},
\end{equation}
where $\beta=\xi\kappa$. For the current analysis, we set
$\kappa=1$, which implies $\beta=\xi$. It is crucial to note that
the one can obtain compact structures which are physically realistic
only for the values of $\beta$ other than $\frac{1}{4}$.

\textcolor{blue} {The presence of anisotropy in the interiors of
compact stars is a fundamental topic in astrophysics, profoundly
affecting their structural and dynamic characteristics. Anisotropy
denotes the variation in pressure within a star, specifically where
radial pressure does not equal the tangential one. This phenomenon
becomes especially significant in high-density environments, such as
neutron stars, where extreme gravitational and magnetic forces can
cause notable departures from isotropic profile. Research on
anisotropic models has progressed since the early 20$^{th}$ century,
with key studies emphasizing the importance of incorporating
anisotropic pressure into stellar modeling. Various investigations
have established that anisotropic pressures can emerge due to
several factors (such as rotation, magnetic fields, and phase
transitions in stellar matter \cite{37n}-\cite{37r})}. This can be
written as \cite{37z1}
\begin{equation}\label{g5}
T_{\eta\mu}=(\rho+P_t)v_{\eta}v_{\mu}+P_t g_{\eta\mu}-\Pi u_\eta
u_\mu,
\end{equation}
in which $\Pi=P_t-P_r$ is the anisotropy (the difference of
tangential and radial pressure), $\rho$ is the fluid's density,
$v_{\eta}$ being the four-velocity. Also, the four-vector is defined
by $u_{\eta}$.

Our aim is to explore the charge effect on existing stellar
structures obtained through the gravitational decoupling.
\textcolor{blue} {This can only be done if we redefine the matter
part of the field equations by adding the electromagnetic tensor
$E_{\eta\mu}$ (already defined in the literature) and additional
fluid source $\Theta_{\eta\mu}$. Also, keep in mind that the later
fluid distribution is added along with a constant, named the
decoupling parameter $\zeta$ that controls its influence on the
geometry under discussion.} The field equations now become as
\begin{eqnarray}\label{g5a}
G_{\eta\mu}=\tilde{T}_{\eta\mu},
\end{eqnarray}
where
\begin{eqnarray}\label{g5b}
\tilde{T}_{\eta\mu}=\bar{T}_{\eta\mu}+E_{\eta\mu}+\zeta\Theta_{\eta\mu},
\end{eqnarray}
and the divergence of the above fluid setup yields
$\nabla^\eta\tilde{T}_{\eta\mu}=0$. The second term guarantees the
presence of charge in a geometrical structure and defined as
\cite{37z2}
\begin{equation*}
E_{\eta\mu}=\frac{1}{4\pi}\left[\frac{1}{4}g_{\eta\mu}\mathcal{N}^{\tau\nu}\mathcal{N}_{\tau\nu}
-\mathcal{N}^{\nu}_{\eta}\mathcal{N}_{\nu\eta}\right],
\end{equation*}
where $\mathcal{N}_{\eta\mu}=\phi_{\mu;\eta}-\phi_{\eta;\mu}$ and
$\phi_{\eta}=\phi(r)\delta^{0}_{\eta}$ are the Maxwell field tensor
and four-potential, respectively.

A spherical geometry either static or non-static has a great
importance in the literature as several researchers believe that the
spacetime structures, such as stars, galaxies, etc. are exactly or
approximately spherically symmetric. Following this, the interior
spherical geometry is defined as
\begin{equation}\label{g6}
ds_-^2=-e^{\sigma} dt^2+e^{\alpha}
dr^2+r^2\big(d\theta^2+\sin^2\theta d\phi^2\big),
\end{equation}
with $\alpha=\alpha(r)$ and $\sigma=\sigma(r)$, showing that the
above metric do not admit change over time. This metric helps in
defining some already expressed tensor quantities in the following
way
\begin{equation}\label{g7}
v^\eta=\delta^\eta_0e^{\frac{-\sigma}{2}}, \quad
u^\eta=\delta^\eta_1e^{\frac{-\alpha}{2}}.
\end{equation}
They are found to agree with certain relations such as $u^\eta
v_{\eta}=0,~u^\eta u_{\eta}=1$ and $v^\eta v_{\eta}=-1$.

By taking Eqs.\eqref{g5a}, \eqref{g5b} and \eqref{g6} into account
simultaneously, we get the gravitational equations as
\begin{align}\label{g8}
&e^{-\alpha}\left(\frac{\alpha'}{r}-\frac{1}{r^2}\right)
+\frac{1}{r^2}=\rho-\zeta\Theta_{0}^{0}+\frac{q^2}{r^4}+\frac{\beta}{4\beta-1}\left(P_r+2P_t-\rho\right),\\\label{g9}
&e^{-\alpha}\left(\frac{1}{r^2}+\frac{\sigma'}{r}\right)
-\frac{1}{r^2}=P_r+\zeta\Theta_{1}^{1}-\frac{q^2}{r^4}-\frac{\beta}{4\beta-1}\left(P_r+2P_t-\rho\right),
\\\label{g10}
&\frac{e^{-\alpha}}{4}\left[\sigma'^2-\alpha'\sigma'+2\sigma''-\frac{2\alpha'}{r}+\frac{2\sigma'}{r}\right]
=P_t+\zeta\Theta_{2}^{2}+\frac{q^2}{r^4}-\frac{\beta}{4\beta-1}\left(P_r+2P_t-\rho\right),
\end{align}
where in the above expressions, prime means
$\frac{\partial}{\partial r}$. Also, the terms accompanied by the
factor involving $\beta$ represent the Rastall effect. By employing
the metric specified in Eq.\eqref{g6}, we can expand the
conservation equation, which must involve additional terms depending
on the considered specific context as
\begin{align}\nonumber
&\frac{dP_r}{dr}+\frac{\sigma'}{2}\left(\rho+P_r\right)+\frac{\zeta\sigma'}{2}
\left(\Theta_{1}^{1}-\Theta_{0}^{0}\right)+\frac{2}{r}\left(P_r-P_t\right)\\\label{g12}
&+\zeta\frac{d\Theta_{1}^{1}}{dr}+\frac{2\zeta}{r}\left(\Theta_{1}^{1}
-\Theta_{2}^{2}\right)-\frac{qq'}{r^4}=\frac{\beta}{1-4\beta}\big(\rho'-P_r'-2P_t'\big).
\end{align}
The significance of Eq.\eqref{g12} lies in its utility for
investigating evolutionary transformations within self-gravitating
systems. It is noteworthy that the incorporation of an additional
source term and charge in the initial anisotropic setup introduces
more complications, allowing for a more comprehensive description of
the system's dynamics. Equivalently, we can say that the number of
unknowns are increased, specifically to
$(\sigma$,$\alpha$,$\rho$,$P_r$,$P_t$,$q$,$\Theta_{0}^{0}$,$\Theta_{1}^{1}$,$\Theta_{2}^{2})$.
As a result, additional constraints must be applied to ensure the
solvability.

\section{Application of Minimal Geometric Deformation}

Recall that the extra source is gravitationally coupled to the
initial analogue, and could be the scalar/vector/tensor field. It
must be stressed here that finding the solution to the gravitational
equations containing multiple physical factors is highly difficult
due to their non-linear nature. \textcolor{blue} {This complication
can be reduced if we utilize a well-known and systematic technique
which has been suggested by Herrera \cite{33} in recent years so
that the under-determined system \eqref{g8}-\eqref{g10} can be
handled. This is referred to the gravitational decoupling as the
equations are decoupled into different sets in this technique. The
idea behind this strategy is that a new metric must be assumed so
that the previously determined field equations can be transformed
into a new reference frame in which the resulting systems of
equations are easy to solve.} We now execute this by introducing a
new metric of the form
\begin{equation}\label{g15}
ds_-^2=-e^{\gamma(r)}dt^2+\frac{1}{\chi(r)}dr^2+r^2\big(d\theta^2+\sin^2\theta
d\phi^2\big).
\end{equation}
The radial and temporal metric components are of great importance in
this scheme because the transformations are implemented on them.
These transformations are
\begin{equation}\label{g16}
\gamma\rightarrow\sigma=\gamma+\zeta t, \quad \chi\rightarrow
e^{-\alpha}=\chi+\zeta f,
\end{equation}
where
\begin{itemize}
\item $t$ is the temporal deformation function,
\item $f$ is that corresponding to the radial component.
\end{itemize}
This article centers on the application of MGD method, which ensures
that the temporal metric potential remains unchanged. By expressing
the two functions as $t\rightarrow 0,~f\rightarrow d_f$, we can then
combine this with Eq.\eqref{g16} to get
\begin{equation}\label{g17}
\gamma\rightarrow\sigma=\gamma, \quad \chi\rightarrow
e^{-\alpha}=\chi+\zeta d_f,
\end{equation}
where $d_f=d_f(r)$. \textcolor{blue} {The fascinating aspect of
these transformations is that they are helpful for maintaining the
spherical object's symmetry.} By applying them to
Eqs.\eqref{g8}-\eqref{g10}, we obtain two distinct systems. The
first system arises when $\zeta=0$, which effectively eliminates the
influence of the newly introduced matter setup, as indicated by the
following expression
\begin{align}\label{g18}
&e^{-\alpha}\left(\frac{\alpha'}{r}-\frac{1}{r^2}\right)
+\frac{1}{r^2}=\rho+\frac{q^2}{r^4}+\frac{\beta}{4\beta-1}\left(P_r+2P_t-\rho\right),\\\label{g19}
&e^{-\alpha}\left(\frac{1}{r^2}+\frac{\sigma'}{r}\right)
-\frac{1}{r^2}=P_r-\frac{q^2}{r^4}-\frac{\beta}{4\beta-1}\left(P_r+2P_t-\rho\right),
\\\label{g20}
&\frac{e^{-\alpha}}{4}\left[\sigma'^2-\alpha'\sigma'+2\sigma''-\frac{2\alpha'}{r}+\frac{2\sigma'}{r}\right]
=P_t+\frac{q^2}{r^4}-\frac{\beta}{4\beta-1}\left(P_r+2P_t-\rho\right),
\end{align}
leads to explicit expressions for fluid triplet
\begin{align}\nonumber
\rho&=\frac{e^{-\alpha}}{2r^4}\bigg[r^2 \big\{2 \big(\beta  r^2
\sigma ''+(r-2 \beta  r) \alpha '+(2 \beta -1) \big(1-e^{\alpha
}\big)\big)\\\label{g18a} &+\beta  r^2 \sigma '^2+\beta  r \sigma '
\big(4-r \alpha '\big)\big\}-2 q^2 e^{\alpha }\bigg],\\\nonumber
P_r&=\frac{e^{-\alpha}}{2r^4}\bigg[2 q^2 e^{\alpha }-r^2 \big\{2
\big(\beta  r^2 \sigma ''-2 \beta r \alpha '+(2 \beta -1)
\big(1-e^{\alpha }\big)\big)\\\label{g19a} &+\beta r^2 \sigma '^2-r
\sigma ' \big(-4 \beta +\beta  r \alpha
'+2\big)\big\}\bigg],\\\nonumber
P_t&=\frac{-e^{-\alpha}}{4r^4}\bigg[r^2 \big\{8 \beta +4 \beta r^2
\sigma ''+(2 \beta -1) r^2 \sigma '^2-2 r^2 \sigma ''+r \sigma
'\\\label{g20a} &\times\big(8 \beta +(r-2 \beta r) \alpha
'-2\big)+(2-8 \beta ) r \alpha '-8 \beta e^{\alpha }\big\}+4 q^2
e^{\alpha }\bigg].
\end{align}

On the other hand, the system which includes the effect of a new
source only is achieved as
\begin{align}\label{g21}
&\Theta_{0}^{0}=\frac{{d_f}'}{r}+\frac{d_f}{r^2},\\\label{g22}
&\Theta_{1}^{1}=d_f\left(\frac{\sigma'}{r}+\frac{1}{r^2}\right),\\\label{g23}
&\Theta_{2}^{2}=\frac{d_f}{4}\left(2\sigma''+\sigma'^2+\frac{2\sigma'}{r}\right)
+{d_f}'\left(\frac{\sigma'}{4}+\frac{1}{2r}\right).
\end{align}
\textcolor{blue} {The MGD approach is notable for its ability to
decouple fluid sources, thereby preventing energy exchange between
them. This decoupling allows the sources to be conserved
individually.} We can now proceed to find the solutions for each of
the resulting systems one by one. We observe six unknowns
($\rho$,$P_r$,$P_t$,$\sigma$,$\alpha$,$q$) in
Eqs.\eqref{g18a}-\eqref{g20a}, leading us to adopt an ansatz from
the existing literature. Additionally, four unknowns ($d_f$,
$\Theta_{0}^{0}$, $\Theta_{1}^{1}$, $\Theta_{2}^{2}$) are present in
Eqs.\eqref{g21}-\eqref{g23}. To proceed, we shall impose constraints
on the $\Theta$-sector in the following sections. At this point, it
is crucial to specify the effective fluid triplet as
\begin{equation}\label{g13}
\tilde{\rho}=\rho-\zeta\Theta_{0}^{0},\quad
\tilde{P}_{r}=P_r+\zeta\Theta_{1}^{1}, \quad
\tilde{P}_{t}=P_t+\zeta\Theta_{2}^{2}.
\end{equation}
The following anisotropy aligns with the total fluid configuration,
and we express it as
\begin{equation}\label{g14}
\tilde{\Pi}=\tilde{P}_{t}-\tilde{P}_{r}=\Pi+\Pi_{\Theta},
\end{equation}
where $\Pi_{\Theta}=\zeta(\Theta_{2}^{2}-\Theta_{1}^{1})$
corresponds to the new source.

\section{Shift from Anisotropic Matter Distribution to Isotropic Analogue}

Assuming that the anisotropy $\tilde{\Pi}$ given by Eq.\eqref{g14}
corresponds to the both matter distributions within the interior
fluid, we shall now investigate the consequences on the massive
structure if this factor deviates from the one corresponding solely
to the initially assumed source. In light of the preceding
discussion, we introduce a constraint that assumes the seed source
becomes isotropic when another source is added, governed by the
parameter $\zeta$. It is essential to note that this assumption is a
mathematical construct that is valid only for a specific value of
this parameter. The constraint is given below as
\begin{equation}\label{g14a}
\Pi_{\Theta}=-\Pi \quad \Rightarrow \quad
\Theta_{1}^{1}-\Theta_{2}^{2}=P_t-P_r.
\end{equation}
Recently, Casadio et al. \cite{37k} utilized this approach in their
analysis on decoupled models, where they transformed the anisotropic
sphere into its isotropic counterpart. Since we need to tackle both
sets of equations individually, we adopt $q(r)=\omega r^3$ with
$\omega$ has a dimension of $\frac{1}{\ell^2}$ \cite{42a11} along
with the Krori-Barua spacetime to handle the first system as
\cite{42a1}
\begin{align}\label{g33}
\sigma(r)&=a_1r^2+a_2,\\\label{g34}
\chi(r)&=e^{-\alpha}=e^{-a_3r^2},\\\nonumber
\rho&=\frac{e^{-a_3r^2}}{r^2}\big[2 \beta  a_1^2 r^4+2 (1-2 \beta )
a_3 r^2-2 \beta  a_1 r^2 \big(a_3 r^2-3\big)\\\label{g35} &-2 \beta
e^{a_3 r^2}+e^{a_3 r^2}-r^4 \omega ^2 e^{a_3 r^2}+2 \beta
-1\big],\\\nonumber P_r&=\frac{e^{-a_3r^2}}{r^2}\big[4 \beta a_3
r^2-2 \beta a_1^2 r^4+2 a_1 r^2 \big(\beta  a_3 r^2-3 \beta
+1\big)+1\\\label{g36} &+2 \beta e^{a_3 r^2}-e^{a_3 r^2}+r^4 \omega
^2 e^{a_3 r^2}-2 \beta \big],\\\nonumber
P_t&=\frac{e^{-a_3r^2}}{r^2}\big[(1-2 \beta ) a_1^2 r^4+(4 \beta -1)
a_3 r^2-r^4 \omega ^2 e^{a_3 r^2}-2 \beta\\\label{g37} &+2 \beta
e^{a_3 r^2}+a_1 r^2 \big\{(2 \beta -1) a_3 r^2-6 \beta
+2\big\}\big],
\end{align}
containing a set of three constants $(a_1,a_2,a_3)$ which are
unknowns at this point. It is necessary to make them known as they
shall be used in the graphical analysis of our developed models in
subsequent sections. The importance of the above ansatz can be
assessed from the aspect that a large body of literature considers
this metric to explore physically relevant interior distributions
both in GR and its modifications \cite{42a2,42a3}.

We now require to perform some calculations at the spherical
interface to make the above three constants known. \textcolor{blue}
{This analysis on the hypersurface refers to the Darmois junction
conditions in which both interior and vacuum exterior spacetimes
meet at a specific point, named the boundary. The Darmois junction
conditions can be categorized into two fundamental forms: the first
involves the continuity of the intrinsic geometry across the
junction, ensuring that the metric tensor is continuous \cite{42a5}.
The second form addresses the continuity of the extrinsic curvature,
which relates to how the two regions bend into each other. This also
ensures that the radial pressure falls to zero at finite value of
$r$, which is known as the star's radius. The exterior metric should
match with the one considered as the interior region in terms of the
properties such as charge, static in nature, etc.} We eventually
adopt the Reissner-Nordstr\"{o}m line element under its mass
$\tilde{M}$ and charge $\mathrm{Q}$ as
\begin{equation}\label{g25}
ds_+^2=-\left(1-\frac{2\tilde{M}}{r}+\frac{\mathrm{Q}^2}{r^2}\right)dt^2
+\left(1-\frac{2\tilde{M}}{r}+\frac{\mathrm{Q}^2}{r^2}\right)^{-1}dr^2+
r^2\big(d\theta^2+\sin^2\theta d\phi^2\big).
\end{equation}
\textcolor{blue} {Using the condition of continuity of metric
components of geometries \eqref{g6} and \eqref{g25} across the
spherical interface provides the following constraints
\begin{equation}\nonumber
g_{tt}^-~{_=^\Sigma}~g_{tt}^+, \quad g_{rr}^-~{_=^\Sigma}~g_{rr}^+,
\quad g_{tt,r}^-~{_=^\Sigma}~g_{tt,r}^+.
\end{equation}}
Substituting the values in above expressions, we get three equations
given by
\begin{align}\label{36}
e^{a_1\mathrm{R}^2+a_2}&=1-\frac{2\tilde{M}}{\mathrm{R}}+\frac{\mathrm{Q}^2}{\mathrm{R}^2},\\\label{36a}
e^{a_3\mathrm{R}^2}&=\left(1-\frac{2\tilde{M}}{\mathrm{R}}+\frac{\mathrm{Q}^2}{\mathrm{R}^2}\right)^{-1},\\\label{36b}
a_1\mathrm{R}e^{a_1\mathrm{R}^2+a_2}&=\frac{\tilde{M}}{\mathrm{R}^2}-\frac{\mathrm{Q}^2}{\mathrm{R}^3},
\end{align}
whose simultaneous solution is
\begin{eqnarray}\label{g37}
a_1&=&\frac{\tilde{M}\mathrm{R}-\mathrm{Q}^2}{\mathrm{R}^2
\big(\mathrm{R}^2-2 \tilde{M}
\mathrm{R}+\mathrm{Q}^2\big)},\\\label{g37a} a_2&=&\ln
\bigg(1-\frac{2\tilde{M}}{\mathrm{R}}+\frac{\mathrm{Q}^2}{\mathrm{R}^2}\bigg)
-\frac{\tilde{M}\mathrm{R}-\mathrm{Q}^2}{\mathrm{R}^2-2\tilde{M}\mathrm{R}+\mathrm{Q}^2},\\\label{g38}
a_3&=&-\frac{1}{\mathrm{R}^2}\ln \bigg(1-\frac{2
\tilde{M}}{\mathrm{R}}+\frac{\mathrm{Q}^2}{\mathrm{R}^2}\bigg).
\end{eqnarray}
\textcolor{blue} {Here, $a_1$ and $a_3$ have dimension of
$\frac{1}{\ell^2}$ whereas $a_2$ is dimensionless.} As we can from
Eqs.\eqref{g37}-\eqref{g38}, the constants are evaluated in term of
experimental data of a compact star model such as its mass and
radius. We, therefore, consider $4U 1820-30$ along with
$\tilde{M}=1.58 \pm 0.06 M_{\bigodot}$ and $\mathrm{R}=9.1 \pm 0.4
km$ \cite{42aa}. Further, the term $\mathrm{Q}$ shall take different
values in graphical analysis to check its impact on the interior
fluid distribution. Merging gravitational equations with
\eqref{g14a}-\eqref{g34} results in
\begin{align}\nonumber
&a_1 r^3 \big\{{d_f}'(r)-2 a_3 r e^{-a_3 r^2}\big\}+2 \big(a_1^2
r^4-1\big) d_f(r)-2 a_3 r^2 e^{-a_3 r^2}\\\label{g39} &-2
e^{-a_3r^2}+2 a_1^2 r^4 e^{-a_3 r^2}+r {d_f}'(r)-4 r^4 \omega
^2+2=0,
\end{align}
whose analytical solution after performing integration (and thus a
constant $c_1$ is appeared) becomes
\begin{align}\nonumber
d_f(r)&=\frac{1}{e^{1+a_1r^2}a_1} \big[2 \omega ^2r^2
\text{Ei}\big(a_1 r^2+1\big)-a_1^2r^2 \text{Ei}\big(a_1
r^2+1\big)\\\label{g40}&+ea_1\big\{e^{a_1 r^2}-e^{(a_1-a_3)
r^2}\big\}\big]+c_1 r^2 e^{-a_1r^2}.
\end{align}
When we merge the above equation with \eqref{g17}, the $g_{rr}$
potential is obtained as
\begin{align}\nonumber
e^{\alpha}=\chi^{-1}&=\bigg[e^{-a_3r^2}+\frac{\zeta}{e^{1+a_1r^2}a_1}
\big\{2 \omega ^2r^2 \text{Ei}\big(a_1 r^2+1\big)-a_1^2r^2
\text{Ei}\big(a_1 r^2+1\big)\\\label{g40a}&+ea_1\big(e^{a_1
r^2}-e^{(a_1-a_3) r^2}\big)\big\}+c_1 \zeta r^2
e^{-a_1r^2}\bigg]^{-1}.
\end{align}

Hence, the solution to the Rastall equations of motion, achieved
through the MGD approach, is represented by the following spacetime
metric
\begin{align}\nonumber
ds_-^2&=-e^{a_1r^2+a_2}dt^2+\bigg[e^{-a_3r^2}+\frac{\zeta}{e^{1+a_1r^2}a_1}
\big\{2 \omega ^2r^2 \text{Ei}\big(a_1
r^2+1\big)\\\nonumber&-a_1^2r^2 \text{Ei}\big(a_1
r^2+1\big)+ea_1\big(e^{a_1 r^2}-e^{(a_1-a_3) r^2}\big)\big\}+c_1
\zeta r^2 e^{-a_1r^2}\bigg]^{-1}dr^2
\\\label{g41} &+ r^2\big(d\theta^2+\sin^2\theta d\phi^2\big).
\end{align}
This metric describes the interior configuration of the fluid, which
is determined by the following fluid variables as
\begin{align}\nonumber
\tilde{\rho}&=a_1 \big\{2 c_1 \zeta  r^2 e^{-a_1 r^2}+3 \zeta
e^{-a_1 r^2-1} \text{Ei}\big(a_1 r^2+1\big)-2 \beta  a_3 r^2 e^{-a_3
r^2}+6 \beta  e^{-a_3 r^2}\big\}\\\nonumber &-3 c_1 \zeta e^{-a_1
r^2}+2 a_1^2 r^2 \bigg\{\beta e^{-a_3 r^2}-\zeta  e^{-a_1 r^2-1}
\text{Ei}\big(a_1 r^2+1\big)+\frac{\zeta }{a_1
r^2+1}\bigg\}+\frac{1}{r^2}\\\nonumber &+4 \zeta r^2 \omega ^2
e^{-a_1 r^2-1} \text{Ei}\big(a_1 r^2+1\big)-\frac{6 \zeta \omega ^2
e^{-a_1 r^2-1} \text{Ei}\big(a_1 r^2+1\big)}{a_1}-\frac{4 \zeta r^2
\omega ^2}{a_1 r^2+1}\\\label{g46} &-2 a_3 (2 \beta +\zeta -1)
e^{-a_3r^2}+\frac{2 \beta  e^{-a_3 r^2}}{r^2}+\frac{\zeta
e^{-a_3r^2}}{r^2}-\frac{e^{-a_3 r^2}}{r^2}-\frac{2 \beta
}{r^2}-\frac{\zeta }{r^2}-r^2 \omega ^2,\\\nonumber
\tilde{P}_{r}&=\frac{e^{-a_3r^2}}{r^2}\big[4 \beta  a_3 r^2-2 \beta
a_1^2 r^4+2 a_1 r^2 \big(\beta  a_3 r^2-3 \beta +1\big)+2 \beta
e^{a_3 r^2}-e^{a_3 r^2}+1\\\nonumber &+r^4 \omega ^2 e^{a_3 r^2}-2
\beta \big]+\frac{\zeta e^{-a_1 r^2-1} \big(2 a_1 r^2+1\big)}{a_1
r^2}\big[e a_1 \big\{e^{a_1 r^2}-e^{(a_1-a_3) r^2}+c_1
r^2\big\}\\\label{g47} &+2 r^2 \omega ^2 \text{Ei}\big(a_1
r^2+1\big)-a_1^2 r^2 \text{Ei}\big(a_1 r^2+1\big)\big],\\\nonumber
\tilde{P}_{t}&=\frac{e^{-a_1 r^2-a_3 r^2-1}}{a_1r^2}\big[a_1^2 r^2
\big\{5 c_1 \zeta  r^2 e^{a_3 r^2+1}-\zeta  e^{a_3 r^2}
\text{Ei}\big(a_1 r^2+1\big)+a_3 (2 \beta +\zeta -1)\\\nonumber
&\times  r^2 e^{a_1 r^2+1}-6 \beta e^{a_1 r^2+1}-5 \zeta  e^{a_1
r^2+1}+5 \zeta e^{a_1 r^2+a_3 r^2+1}+2 e^{a_1 r^2+1}\big\}+a_1
\big(c_1 \zeta \\\nonumber &\times r^2 e^{a_3 r^2+1}+10 \zeta  r^4
\omega ^2 e^{a_3 r^2} \text{Ei}\big(a_1 r^2+1\big)+a_3 r^2 (4 \beta
+\zeta -1) e^{a_1 r^2+1}-2 \beta e\\\nonumber &\times e^{a_1 r^2}+2
\beta  e^{a_1 r^2+a_3 r^2+1}+2 \zeta  r^4 \omega ^2 e^{a_1 r^2+a_3
r^2+1}-r^4 \omega ^2 e^{a_1 r^2+a_3 r^2+1}\big)+2 \zeta r^2 \omega
^2 \\\label{g48} &\times e^{a_3 r^2} \text{Ei}\big(a_1
r^2+1\big)-a_1^3 r^4 \big\{5 \zeta e^{a_3 r^2} \text{Ei}\big(a_1
r^2+1\big)+(2 \beta +\zeta -1) e^{a_1 r^2+1}\big\}\big],
\end{align}
and the anisotropic factor becomes in this scenario as
\begin{equation}\label{g49}
\tilde{\Pi}=\frac{e^{-a_3r^2}}{r^2}\big(a_1^2 r^4-a_1 a_3 r^4-a_3
r^2+e^{a_3 r^2}-2 r^4 \omega ^2 e^{a_3 r^2}-1\big)\big(1-\zeta\big).
\end{equation}
We clearly see from the above equation that the substitution of
$\zeta=1$ results in the vanishing anisotropy. Hence, we can say
that the first unique solution \eqref{g46}-\eqref{g49} to the
charged field equations belong to $\zeta\in[0,1]$. It is also needed
to mention that conversion of anisotropic to the isotropic analogue
or vice versa depends on the variation of the decoupling parameter
within the mentioned range.

\subsection{Physical Acceptability Requirements and their Graphical Confirmation}

Some properties are necessary to be checked in the interior fluid
distribution corresponding to the resulting model so that its
physical feasibility can be assessed. We start off from the
spherical mass expressed by the following differential equation
\cite{37z,37za}
\begin{equation}\label{g63}
m'(r)=\frac{1}{2} \tilde{\rho} r^2 \quad \mathrm{or} \quad
m(r)=\frac{1}{2}\int_{0}^{r} \tilde{\rho} \tilde{r}^2 d\tilde{r},
\end{equation}
that can be solved either through exact or numerical integration
depending on the complexities in the overhead equation. In this
case, the effective form of matter's energy density is provided in
Eq.\eqref{g46}. Secondly, a compactness factor must be measured for
our geometry which defines the closeness of the particles in any
celestial body due to its own gravitational pull. Or in other words,
one can express it by the ratio as $\varpi(r)=\frac{m(r)}{r}$ that
must be less than $\frac{4}{9}$ in spherical spacetime. This limit
has been measured in a study done by Hans Adolph Buchdahl \cite{42a}
in 1959. Thirdly, another quantity depends on the compactness factor
and named the surface redshift. This actually measures the
persistent grow (decline) in the wavelength (frequency) of the
radiations emitted by heavily compact bodies when they are
influenced by other nearby massive objects. This factor is
\begin{equation}
\textrm{z}(r)=\frac{1}{\sqrt{1-2\varpi(r)}}-1,
\end{equation}
whose maximum value is classified as
\begin{itemize}
\item For isotropic fluid geometry, it is found as 2 \cite{42a},
\item 5.211 when anisotropy appears in the principal pressure \cite{42b}.
\end{itemize}

In evaluating the viability of a model, astrophysicists have put
forth certain limitations. Adhering to these constraints ensures the
presence of a usual (normal) fluid within the specified interior.
Failure to meet these criteria would necessitate the presence of an
exotic fluid, thereby impeding the physical viability of stellar
structures. These boundaries predominantly rely on the EMT,
expresses in the context of anisotropic geometry as
\begin{equation}
\left.
\begin{aligned}\label{g50}
&\tilde{\rho}+\frac{q^2}{r^4} \geq 0, \quad
\tilde{\rho}+\tilde{P}_{r} \geq 0, \\
&\tilde{\rho}+\tilde{P}_{t}+\frac{2q^2}{r^4} \geq 0, \quad
\tilde{\rho}-\tilde{P}_{r}+\frac{2q^2}{r^4} \geq 0,\\
&\tilde{\rho}-\tilde{P}_{t} \geq 0, \quad
\tilde{\rho}+\tilde{P}_{r}+2\tilde{P}_{t}+\frac{2q^2}{r^4} \geq 0.
\end{aligned}
\right\}
\end{equation}
Rather than examining all above constraints, one can check only
dominant bounds expressed by
$\tilde{\rho}-\tilde{P}_{r}+\frac{2q^2}{r^4} \geq 0$ and
$\tilde{\rho}-\tilde{P}_{t} \geq 0$ when all fluid variables lie in
the positive range.

Stability is the most important property to be checked when studying
stellar models. Here, we shall use two criteria depend on the sound
speed discussed below
\begin{itemize}
\item It has been proposed in the literature that if the sound speed
in both radial ($\tilde{v}_r$) and tangential ($\tilde{v}_t$)
directions are less of the light's speed, then the model is
considered as stable \cite{42c}. We can express it as
\begin{equation}\label{g62b}
0 < \tilde{v}_r=\frac{d\tilde{P}_{r}}{d\tilde{\rho}} < 1, \quad 0 <
\tilde{v}_t=\frac{d\tilde{P}_{t}}{d\tilde{\rho}} < 1.
\end{equation}
\item Herrera \cite{42ba} put forwarded the idea of a
cracking in the interior distribution. According to his study, if
the inequality
\begin{equation}\label{g62a}
0 < |\tilde{v}_t-\tilde{v}_r| < 1,
\end{equation}
satisfies, the system remains stable. Otherwise, it becomes unstable
and no more physically relevant.
\end{itemize}

Now, we explore all the above-mentioned characteristics for our
developed model \eqref{g46}-\eqref{g49} by varying the parameters
involved in these mathematical results. We adopt two, ten and two
distinct values of the Rastall parameter $\beta$, decoupling
parameter $\zeta$ and charge, respectively. More specifically, we
choose $\beta=0.2,0.6$, $\zeta=0.1,0.2,...,1$, $\mathrm{Q}=0.2,0.7$
and $c_{1}=-0.02$. \textcolor{blue} {For choosing acceptable value
of $c_1$, we perform a detailed analysis with its different (small
and large) values. By doing so, we come to know that only small but
negative value of this constant leads to the meaningful results.}
Both metric potentials must be positively increasing function of the
radial coordinate, however, we only plot the radial component
\eqref{g40a} in Figure \textbf{1} as the modification is seen only
in this component due to the applicability of MGD scheme. The
triplet functions such as density and principal pressures
\eqref{g46}-\eqref{g48} are plotted in Figures \textbf{2} and
\textbf{3} (correspond to lower and upper values of charge),
achieving their maximum at $r=0$ and minimum at the interface. It is
noticed that the density and both pressures are in direct and
inverse relation with $\beta$ and $\mathrm{Q}$. However, increment
in $\zeta$ results in decreasing the energy density and increasing
pressure ingredients. As the plots of anisotropy are concerned, all
of them show nullification of this factor for $\zeta=1$, confirming
the presence of mathematical isotropy. However, for all other
parametric choices, this function increases with the rise in $r$.
\begin{figure}[h!]\center
\epsfig{file=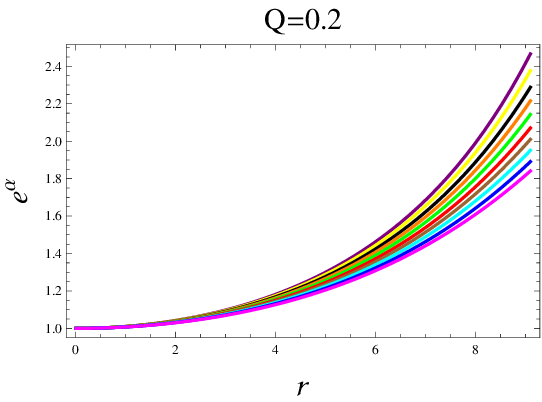,width=.38\linewidth}\epsfig{file=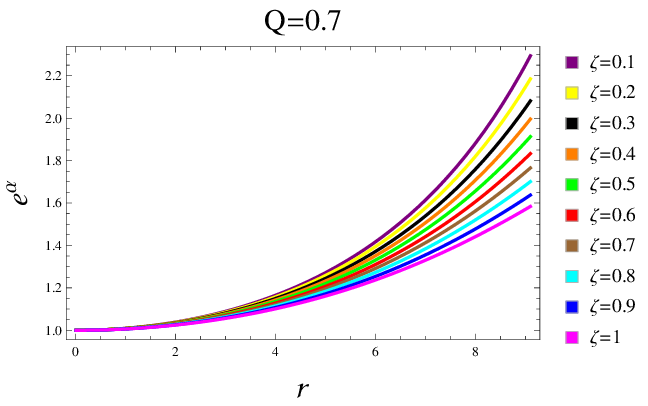,width=.45\linewidth}
\caption{Radial metric \eqref{g40a}.}
\end{figure}
\begin{figure}[h!]\center
\epsfig{file=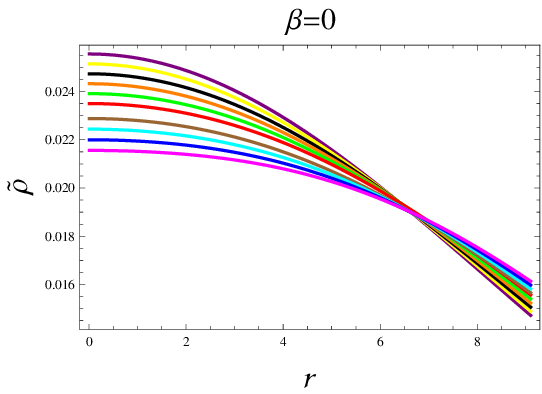,width=.35\linewidth}\epsfig{file=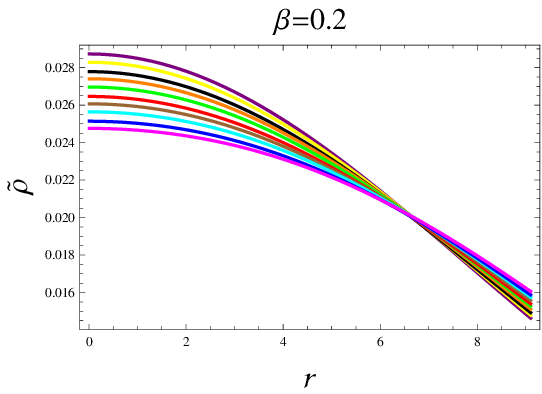,width=.35\linewidth}\epsfig{file=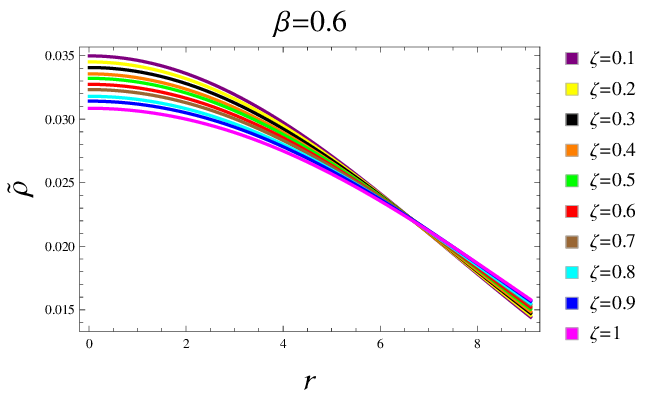,width=.42\linewidth}
\epsfig{file=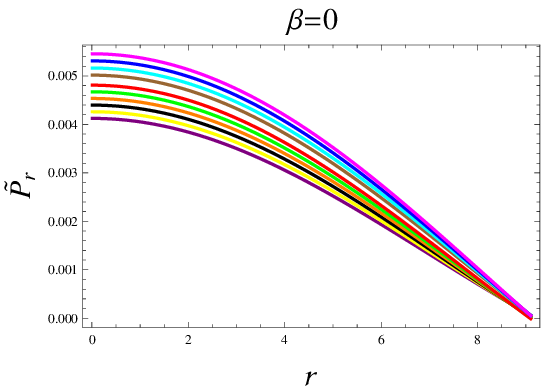,width=.35\linewidth}\epsfig{file=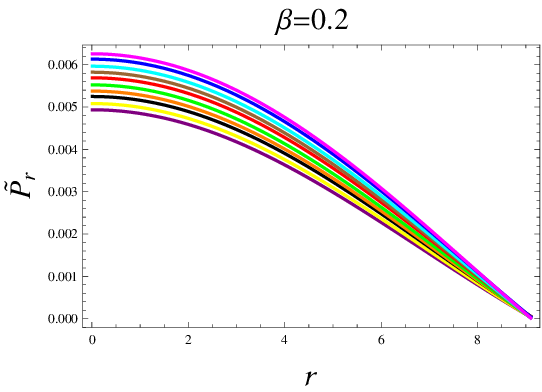,width=.35\linewidth}\epsfig{file=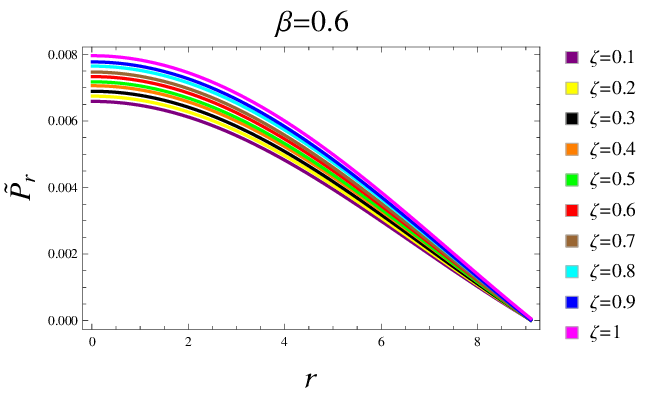,width=.42\linewidth}
\epsfig{file=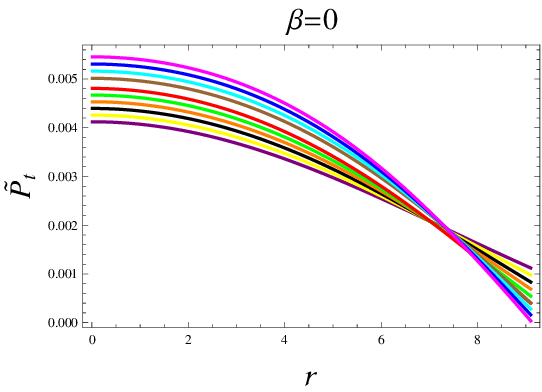,width=.35\linewidth}\epsfig{file=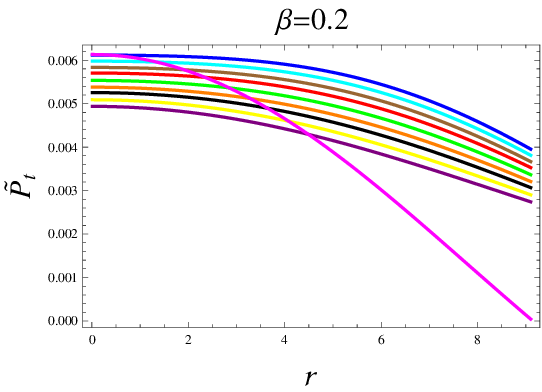,width=.35\linewidth}\epsfig{file=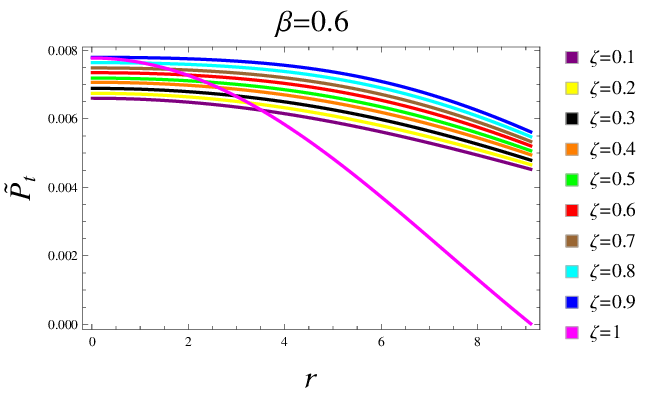,width=.42\linewidth}
\epsfig{file=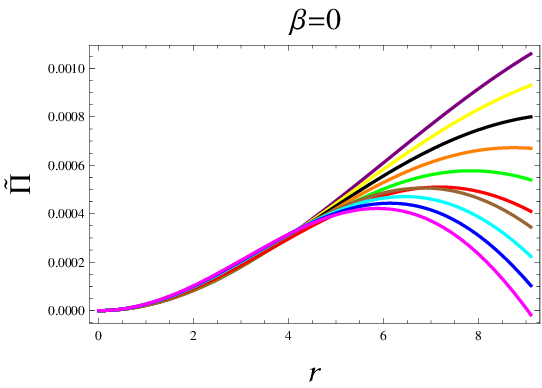,width=.35\linewidth}\epsfig{file=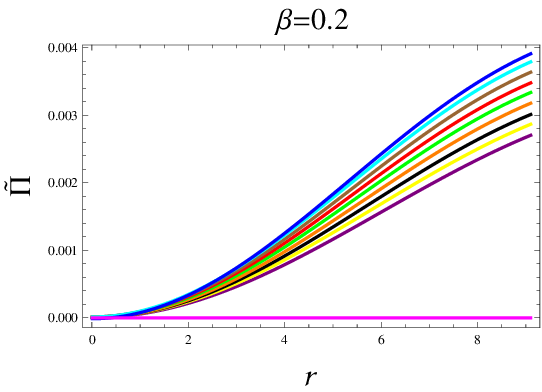,width=.35\linewidth}\epsfig{file=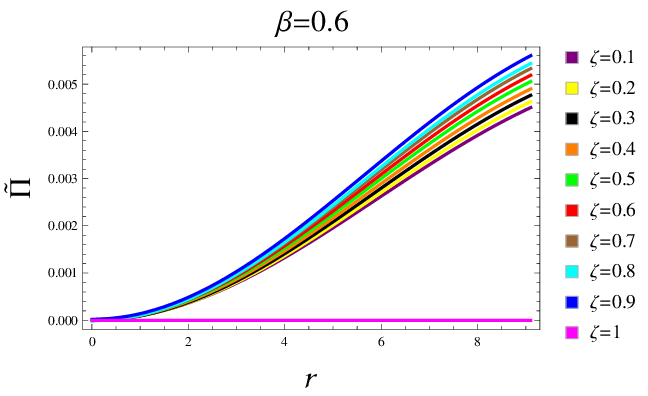,width=.42\linewidth}
\caption{Matter variables corresponding to $\mathrm{Q}=0.2$.}
\end{figure}
\begin{figure}[h!]\center
\epsfig{file=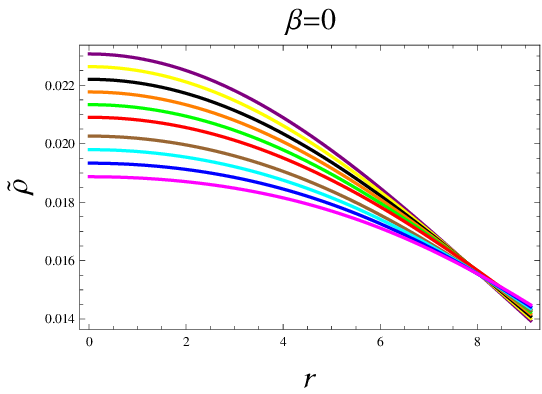,width=.35\linewidth}\epsfig{file=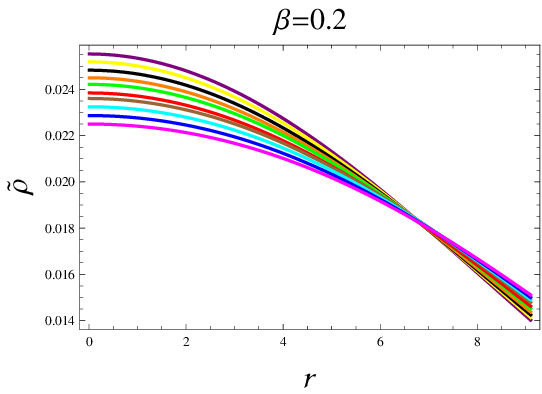,width=.35\linewidth}\epsfig{file=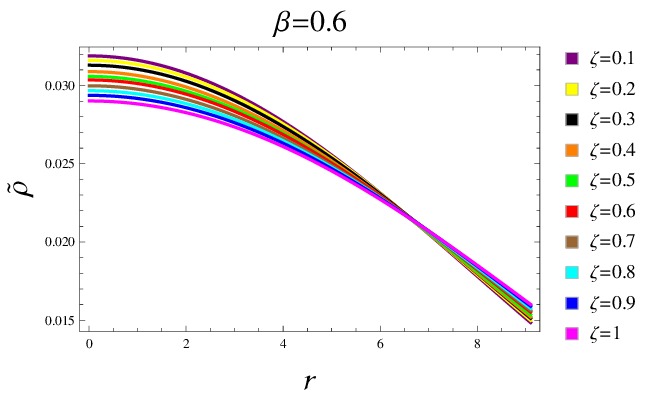,width=.42\linewidth}
\epsfig{file=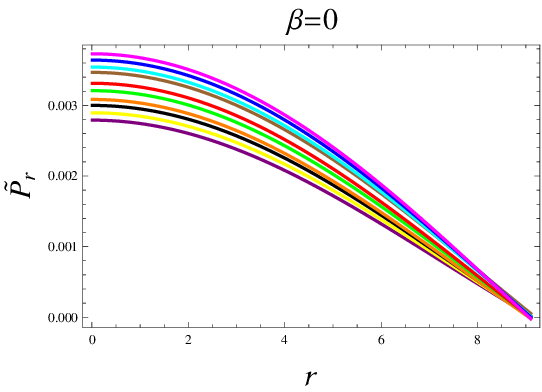,width=.35\linewidth}\epsfig{file=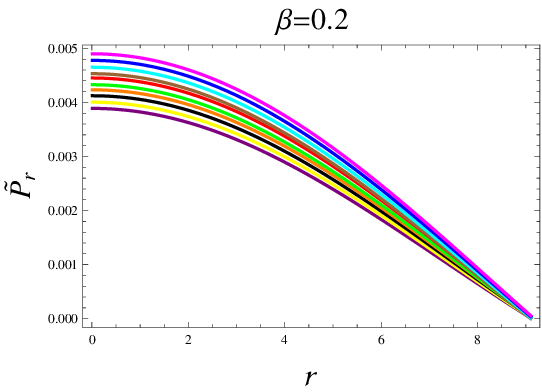,width=.35\linewidth}\epsfig{file=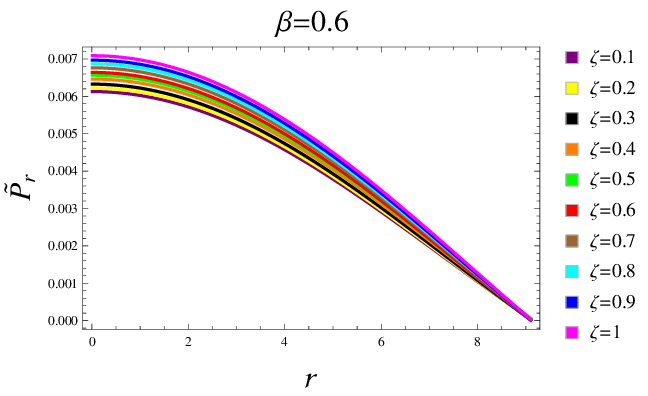,width=.42\linewidth}
\epsfig{file=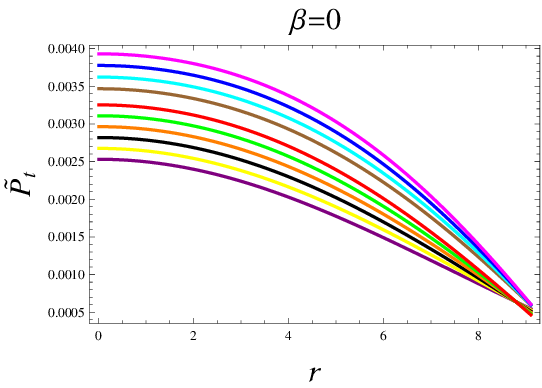,width=.35\linewidth}\epsfig{file=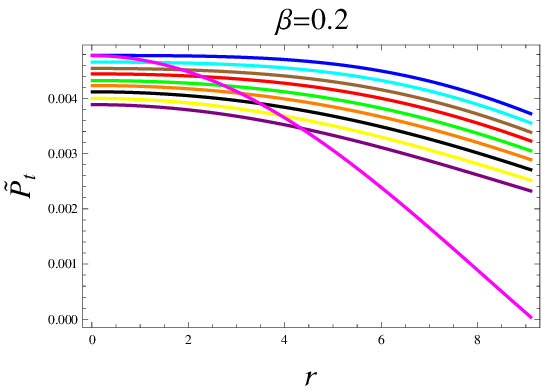,width=.35\linewidth}\epsfig{file=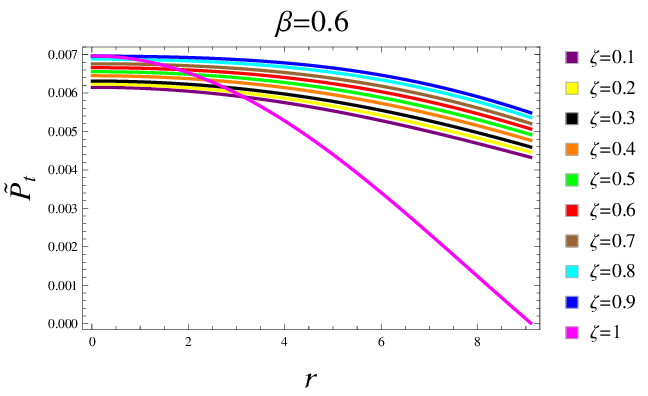,width=.42\linewidth}
\epsfig{file=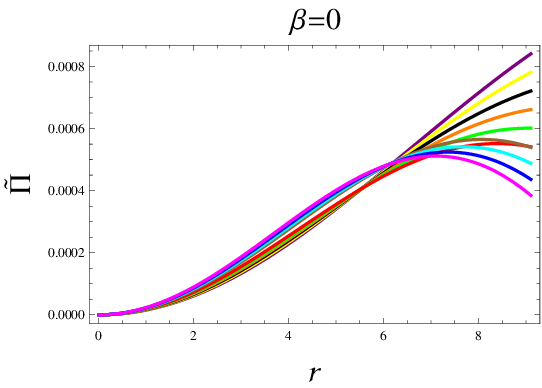,width=.35\linewidth}\epsfig{file=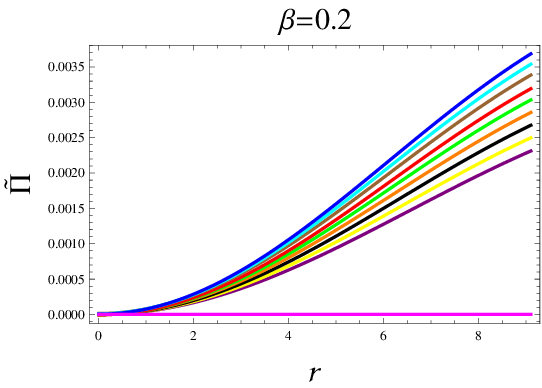,width=.35\linewidth}\epsfig{file=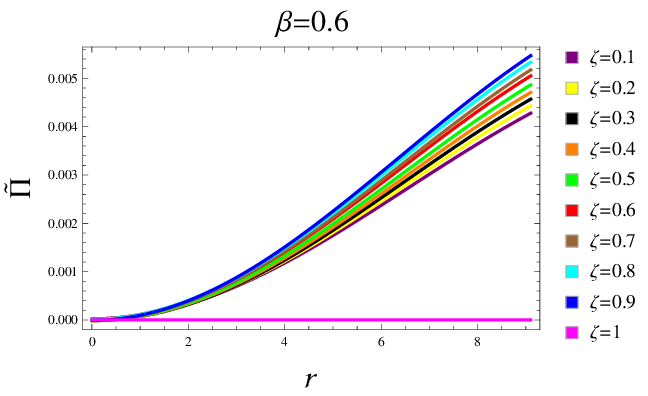,width=.42\linewidth}
\caption{Matter variables corresponding to $\mathrm{Q}=0.7$.}
\end{figure}

Since the mass function depends on the density of fluid (defined in
Eq.\eqref{g63}), we plot this for all parametric variations and find
its increasing profile w.r.t. $r$. \textcolor{blue} {We find the
more dense interior corresponding to this model for $\beta=0.6$,
$\zeta=0.1$ and $\mathrm{Q}=0.2$. The numerical values of the mass
function for the previous parametric choices of $\beta$ and $\zeta$
are $\tilde{M}=1.644M_{\bigodot}$ and $1.609M_{\bigodot}$ for
$Q=0.2$ and $0.7$, respectively. However, the best fit mass with the
observed data is found for lower choices of $\zeta$ and higher
values of the electric charge. We also check other two factors
(redshift and compactness), however, their graphs are not included
here.} The energy conditions admitting the difference between
density and pressures are checked without adding their graphs. All
of them show positive profile of both constraints, hence, we refer
this model to be physically viable. The presence of the usual matter
is also confirmed. Furthermore, both stability criteria are explored
in Figures \textbf{4} and \textbf{5}, respectively. All plots in
both these Figures lie in their desired range. This implies that the
resulting model corresponding to the constraint \eqref{g14a} is
stable.
\begin{figure}[h!]\center
\epsfig{file=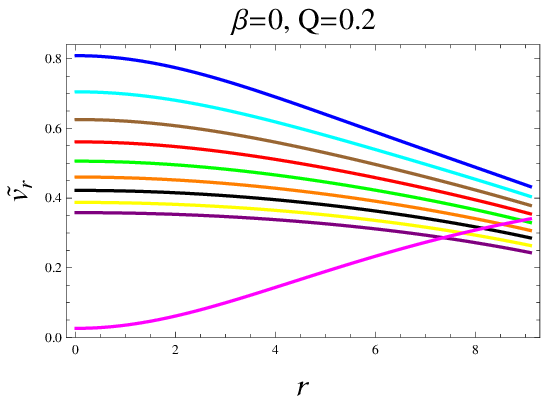,width=.35\linewidth}\epsfig{file=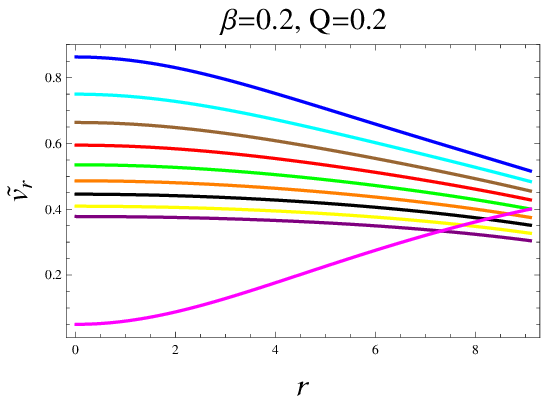,width=.35\linewidth}\epsfig{file=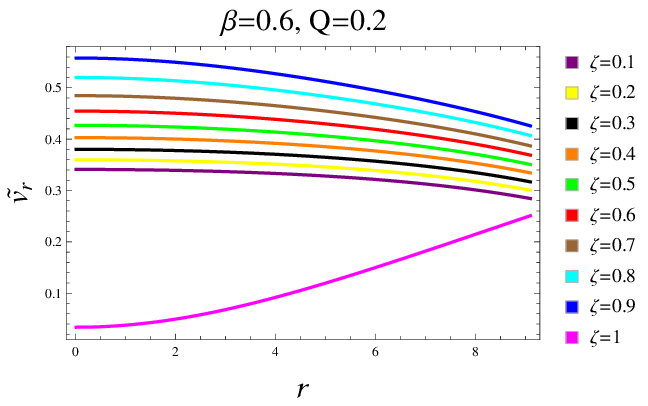,width=.42\linewidth}
\epsfig{file=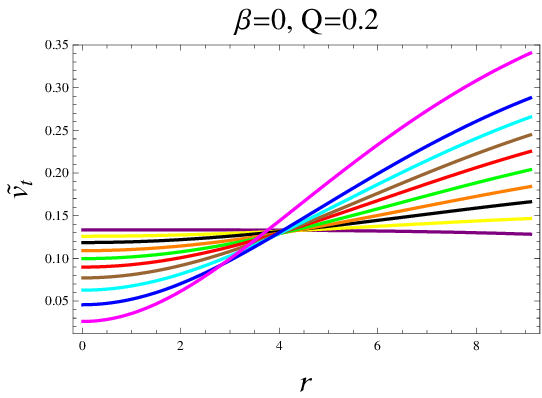,width=.35\linewidth}\epsfig{file=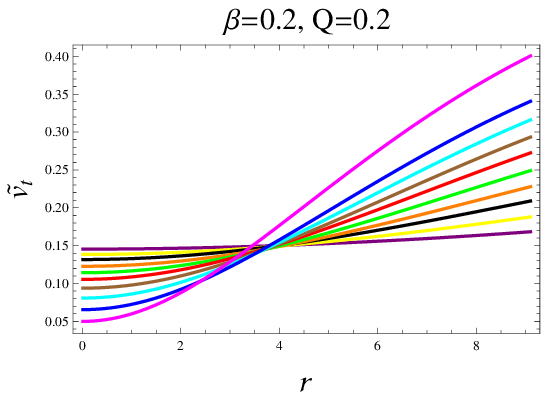,width=.35\linewidth}\epsfig{file=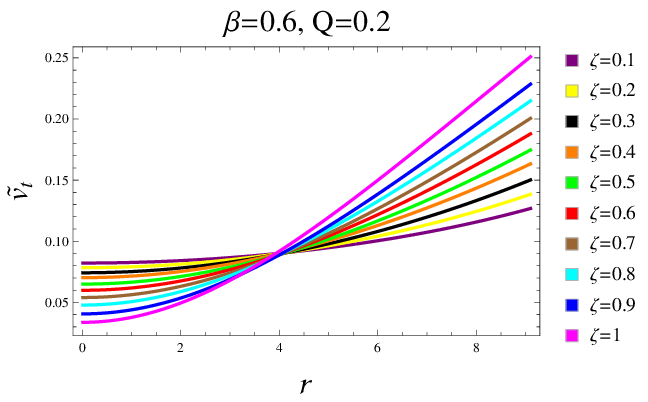,width=.42\linewidth}
\epsfig{file=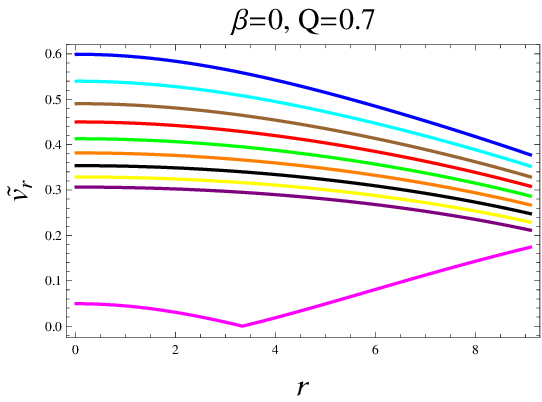,width=.35\linewidth}\epsfig{file=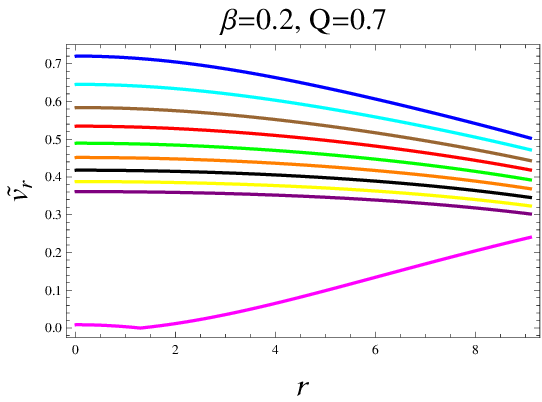,width=.35\linewidth}\epsfig{file=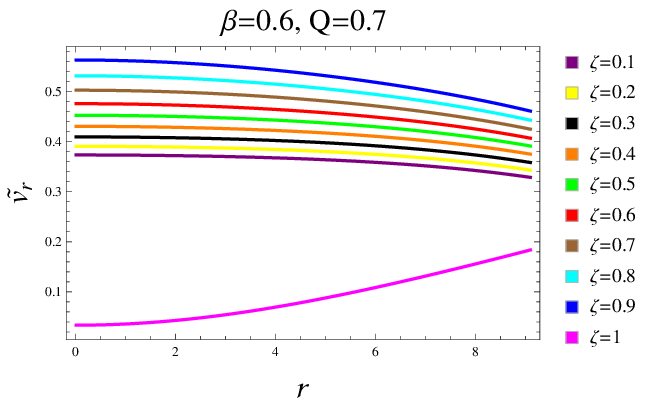,width=.42\linewidth}
\epsfig{file=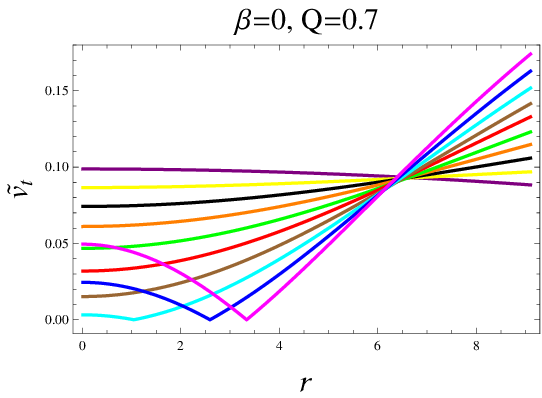,width=.35\linewidth}\epsfig{file=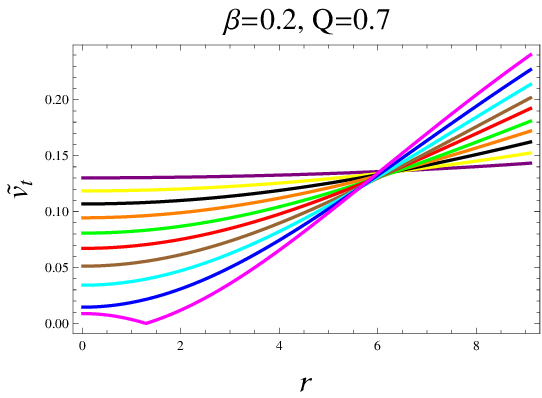,width=.35\linewidth}\epsfig{file=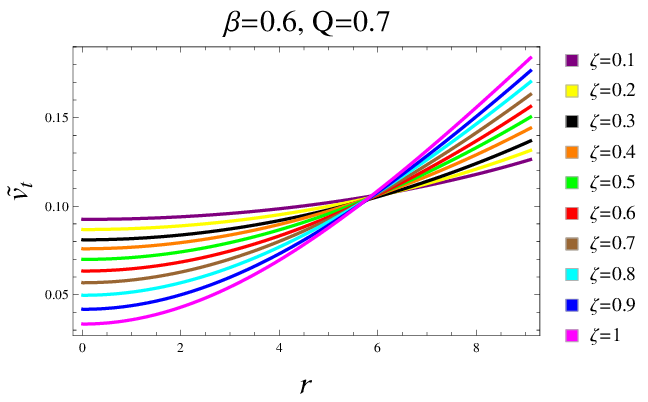,width=.42\linewidth}
\caption{Sound speeds in radial and tangential directions.}
\end{figure}
\begin{figure}[h!]\center
\epsfig{file=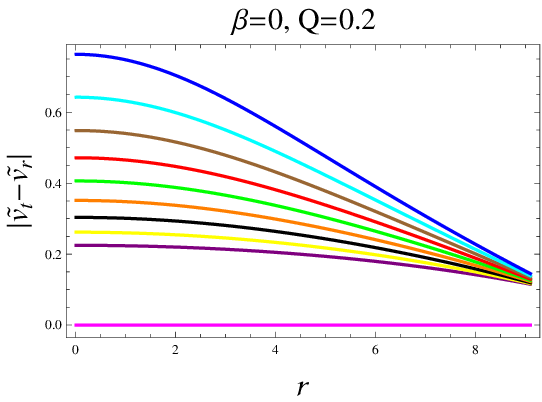,width=.35\linewidth}\epsfig{file=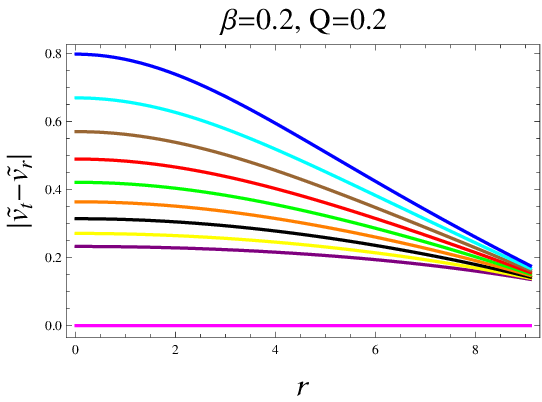,width=.35\linewidth}\epsfig{file=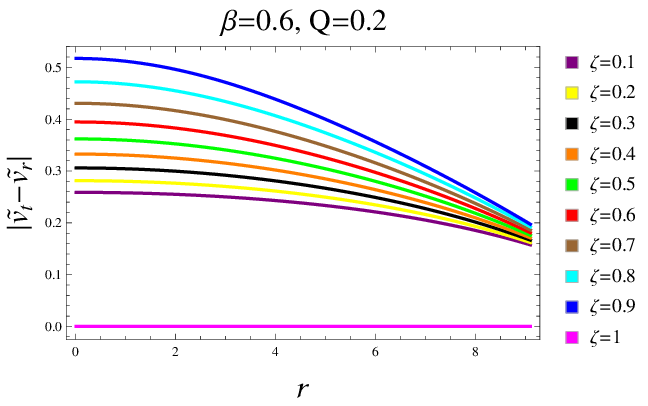,width=.42\linewidth}
\epsfig{file=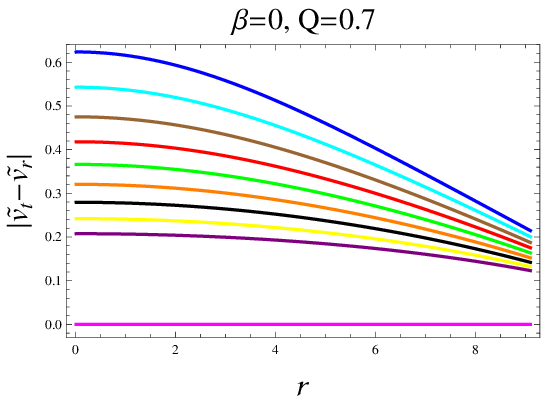,width=.35\linewidth}\epsfig{file=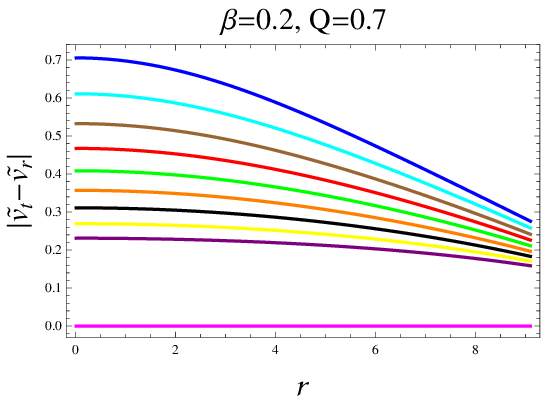,width=.35\linewidth}\epsfig{file=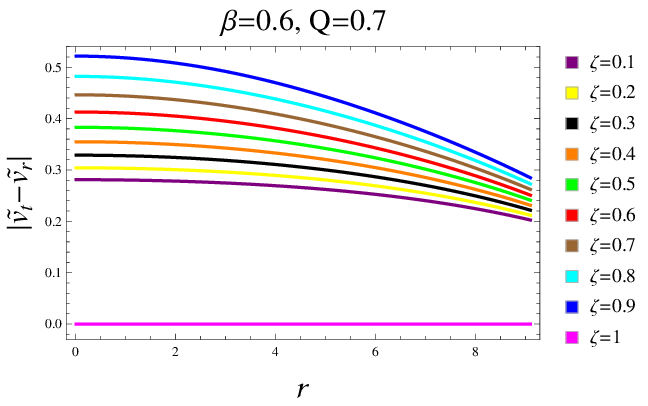,width=.42\linewidth}
\caption{Herrera's cracking criteria.}
\end{figure}

\section{Complexity Factor and its Implication in Compact Matter Source}

The study of astronomical structures has garnered substantial
attention from researchers in recent years. Herrera's work on a
static sphere highlights that the physical factors producing
complexity play a crucial role in the evolution of compact stars,
shedding new light on the intricate dynamics of these celestial
bodies \cite{37g}. The concept was further developed to account for
non-static dynamical fluid setup \cite{37h}. By performing an
orthogonal splitting of the Riemann-Christoffel tensor, several
distinct scalars are identified. It is observed that the two primary
factors influencing fluid content (namely density inhomogeneity and
anisotropic pressure) are incorporated into one of these scalars,
which Herrera referred to as the complexity factor, denoted as
$Y_{TF}$ (for more details, see \cite{42bb}-\cite{42bf}). This is
mathematically represented as
\begin{equation}\label{g51}
Y_{TF}(r)=\Pi+\frac{4q^2}{r^4}-\frac{1}{2r^3}\int_0^r\tilde{r}^3\rho'(\tilde{r})d\tilde{r}
-\frac{3}{r^3}\int_0^r\frac{qq'}{\tilde{r}}d\tilde{r}.
\end{equation}
The above equation denotes the complexity factor for the anisotropic
fluid distribution influenced by an electromagnetic field. Extending
the same definition for our current scenario where two fluid sources
are present, Eq.\eqref{g51} switches into
\begin{eqnarray}\nonumber
\tilde{Y}_{TF}(r)&=&\tilde{\Pi}+\frac{4q^2}{r^4}
-\frac{1}{2r^3}\int_0^r\tilde{r}^3\tilde{\rho}'(\tilde{r})d\tilde{r}
-\frac{3}{r^3}\int_0^r\frac{qq'}{\tilde{r}}d\tilde{r}\\\label{g54}
&=&Y_{TF}+Y_{TF}^{\Theta},
\end{eqnarray}
where
\begin{itemize}
\item $Y_{TF}$ corresponds to the geometry defined by Eqs.\eqref{g18a}-\eqref{g20a},
\item $Y_{TF}^{\Theta}$ being the complexity factor in relation with the newly added
source represented by Eqs.\eqref{g21}-\eqref{g23}.
\end{itemize}
Further, they have the values as follows
\begin{eqnarray}\label{g55}
Y_{TF}&=&\Pi+\frac{4q^2}{r^4}-\frac{1}{2r^3}\int_0^r\tilde{r}^3\rho'(\tilde{r})d\tilde{r}
-\frac{3}{r^3}\int_0^r\frac{qq'}{\tilde{r}}d\tilde{r},\\\label{g55a}
Y_{TF}^{\Theta}&=&\Pi_{\Theta}+\frac{1}{2r^3}\int_0^r\tilde{r}^3\Theta{_0^0}'(\tilde{r})d\tilde{r}.
\end{eqnarray}

Recall that the first resulting solution expressed by
Eqs.\eqref{g46}-\eqref{g49} corresponded to $\tilde{\Pi}=0$
(isotropization of anisotropic model). In view of this,
Eq.\eqref{g54} becomes
\begin{eqnarray}\label{g56}
\tilde{Y}_{TF}&=&\frac{4q^2}{r^4}
-\frac{1}{2r^3}\int_0^r\tilde{r}^3\tilde{\rho}'(\tilde{r})d\tilde{r}
-\frac{3}{r^3}\int_0^r\frac{qq'}{\tilde{r}}d\tilde{r}.
\end{eqnarray}
This along with the fluid's density \eqref{g46} turns into
\begin{align}\nonumber
\tilde{Y}_{TF}&=a_1^2 \bigg\{r^2 \big(\zeta  \tau _1 e^{-a_1
r^2-1}-\beta  e^{-a_3r^2}\big)-\frac{3 \beta e^{-a_3r^2}}{2
a_3}-\frac{9 \beta  e^{-a_3 r^2}}{4 a_3^2 r^2}+\tau
_2\bigg\}\\\nonumber &+a_3 (2 \beta +\zeta -1) e^{-a_3r^2}+\frac{2
\beta e^{-a_3 r^2}}{r^2}-2 \zeta r^2 \tau _1 \omega ^2 e^{-a_1
r^2-1}-\frac{2 \zeta  \omega ^2}{a_1^2 r^2+a_1}\\\nonumber &+a_1
\bigg(\beta a_3 r^2 e^{-a_3r^2}-\frac{3}{2} \beta
e^{-a_3r^2}-\frac{9 \beta e^{-a_3 r^2}}{4 a_3 r^2}-\zeta
r^2c_1e^{-a_1r^2}-\zeta+\tau_2\bigg)\\\label{g56a} &+\frac{\zeta
e^{-a_3r^2}}{r^2}-\frac{e^{-a_3r^2}}{r^2}-\frac{\zeta }{a_1
r^4+r^2}-\frac{2 \beta }{r^2}+\frac{12 r^2 \omega
^2}{5}+\frac{1}{r^2}+\frac{2 \zeta  \omega ^2}{a_1},
\end{align}
where
\begin{align}\nonumber
\tau_1&=\text{Ei}\big(a_1 r^2+1\big),\\\nonumber \tau_2&=\frac{9
\sqrt{\pi } \beta~  \text{erf}\left(\sqrt{a_3} r\right)}{8 a_3^{5/2}
r^3}.
\end{align}
This factor is graphically explored for all parametric choices
(defined earlier) in Figure \textbf{6}. We do not consider the
specific values of the complexity factor; instead, we analyze its
behavior in relation to various parameters. At the core, this factor
reaches a value of zero, indicating that the two factors
contributing to complexity effectively negate each other's influence
at this juncture. Further, we see that this depicts directly related
behavior w.r.t. $r$ and inversely related with $\beta$, $\zeta$ and
charge. \textcolor{blue} {Further, the maximum complexity produced
in the system corresponds to $\zeta=0.1$, $\beta=0.2$ and
$\mathrm{Q}=0.2$. Hence, increasing Rastall parameter results in the
less complexity, showing positive perspective of this extended
theory.}
\begin{figure}[h!]\center
\epsfig{file=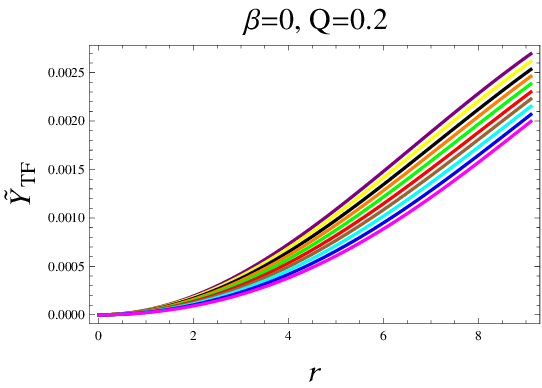,width=.35\linewidth}\epsfig{file=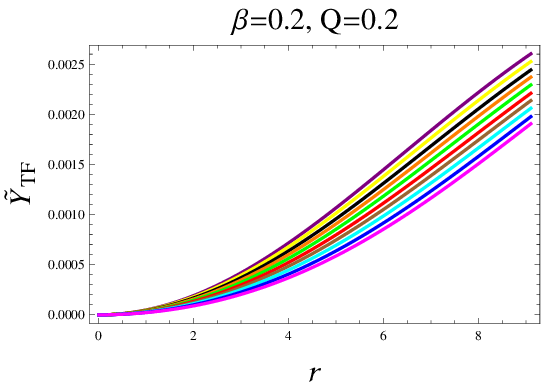,width=.35\linewidth}\epsfig{file=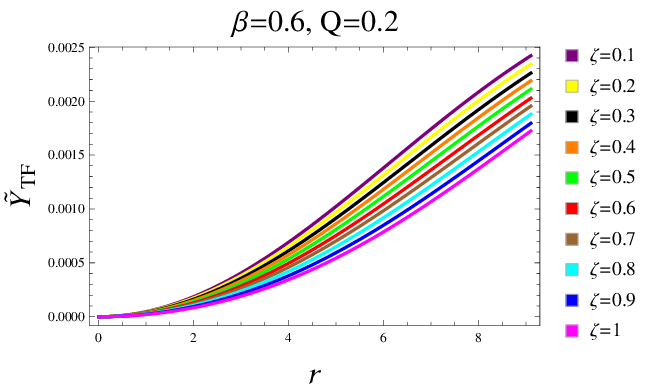,width=.42\linewidth}
\epsfig{file=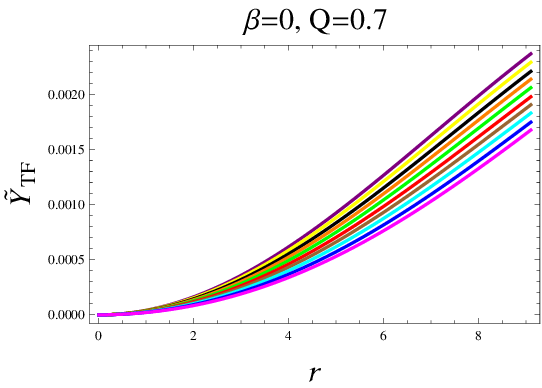,width=.35\linewidth}\epsfig{file=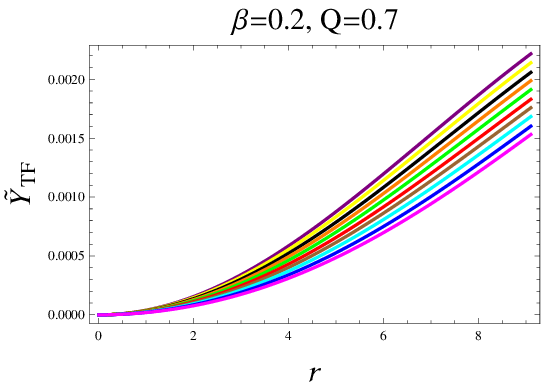,width=.35\linewidth}\epsfig{file=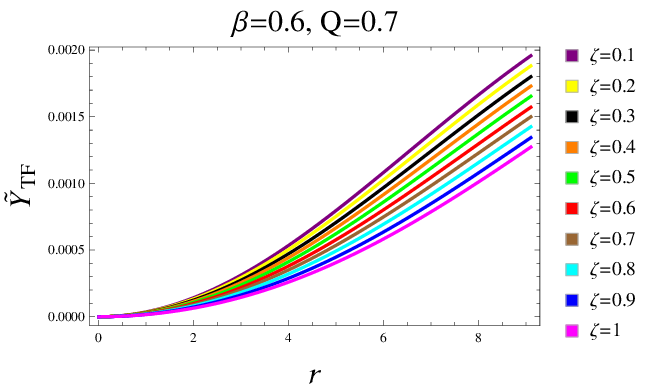,width=.42\linewidth}
\caption{Complexity factor \eqref{g56a}.}
\end{figure}

\subsection{Additional Fluid Setup admitting Null Complexity}

Here, we consider the scenario that adding the new fluid source
makes no change in the complexity of initial source, i.e.,
$Y_{TF}^{\Theta}=0$. In other words, the complexity of initial and
total matter distribution are same. This is mathematically defined
as $\tilde{Y}_{TF}=Y_{TF}$, leads to
\begin{eqnarray}\label{g56b}
\Pi_{\Theta}=-\frac{1}{2r^3}\int_0^r\tilde{r}^3\Theta{_0^0}'(\tilde{r})d\tilde{r}.
\end{eqnarray}
Solving the right side of the above equation by using \eqref{g21},
we obtain
\begin{eqnarray}\label{g56c}
-\frac{1}{2r^3}\int_0^r\tilde{r}^3\Theta{_0^0}'(\tilde{r})d\tilde{r}=\frac{d_f}{r^2}-\frac{{d_f}'}{2r},
\end{eqnarray}
and substituting this back into Eq.\eqref{g56b}, the linear-order
equation becomes in terms of the differential of a function $d_f(r)$
as
\begin{align}\label{g56d}
r {d_f}'(r) \big(r \sigma '+4\big)+d_f(r) \big(2 r^2 \sigma ''+r^2
\sigma '^2-2 r \sigma '-8\big)=0.
\end{align}
This single equation contains two unknowns, one deformation function
and one metric potential. One can solve it for the former factor
only by assuming a known form of the later quantity. We, therefore,
adopt the Tolman IV ansatz to solve Eq.\eqref{g56d} expressed by
\begin{align}\label{g57}
\sigma(r)&=\ln\bigg\{a_5\bigg(1+\frac{r^2}{a_4}\bigg)\bigg\},\\\label{g58}
\chi(r)&=e^{-\alpha}=\frac{\big(a_4+r^2\big)\big(a_6-r^2\big)}{a_6\big(a_4+2r^2\big)},
\end{align}
whose corresponding fluid components are
\begin{align}\nonumber
\rho&=\frac{1}{a_6 \big(a_4+2 r^2\big)^2}\big[a_4^2 \big(3-a_6 r^2
\omega ^2-6 \beta\big)+2 r^2 \big\{a_6 \big(2 \beta -2 r^4 \omega
^2+1\big)\\\label{g59} &+3 (1-4 \beta ) r^2\big\}+a_4 \big\{a_6
\big(3-4 r^4 \omega ^2\big)+(7-22 \beta ) r^2\big\}\big],\\\nonumber
P&=\frac{1}{a_6 \big(a_4+2 r^2\big)^2}\big[a_4^2 \big(a_6 r^2 \omega
^2+6 \beta -1\big)+2 r^2 \big\{a_6 \big(1-2 \beta +2 r^4 \omega
^2\big)\\\label{g60} &+3 (4 \beta -1) r^2\big\}+a_4 \big\{a_6 \big(4
r^4 \omega ^2+1\big)+(22 \beta -5) r^2\big\}\big].
\end{align}
To find the triplet ($a_4,a_5,a_6$) appeared in Eqs.\eqref{g57} and
\eqref{g58}, we need to use the Reissner-Nordstr\"{o}m metric
\eqref{g25} again. Their expressions in the presence of charge take
the form
\begin{align}\label{g60a}
a_4&=\frac{\mathrm{R}^2\big(2\mathrm{R}^2-6\tilde{M}\mathrm{R}+3\mathrm{Q}^2\big)}
{2\tilde{M}\mathrm{R}-\mathrm{Q}^2},\\\label{g60aa}
a_5&=\frac{2\mathrm{R}^2-6\tilde{M}\mathrm{R}+3\mathrm{Q}^2}{2\mathrm{R}^2},\\\label{g60aaa}
a_6&=\frac{2\mathrm{R}^3\big(\mathrm{Q}^2-2\tilde{M}\mathrm{R}+\mathrm{R}^2\big)}
{2\tilde{M}\mathrm{Q}^2-4\tilde{M}^2\mathrm{R}+\mathrm{Q}^2\mathrm{R}+2\tilde{M}\mathrm{R}^2}.
\end{align}
We now incorporate the temporal coefficient \eqref{g57} in
Eq.\eqref{g56d} and perform integration to get $d_f(r)$ as
\begin{equation}\label{g60b}
d_f(r)=\frac{c_2 r^2 \big(a_4+r^2\big)}{2 a_4+3 r^2},
\end{equation}
with an arbitrary constant $c_2$. Joining this with Eq.\eqref{g17}
leads to the modified radial component as
\begin{align}\label{g60c}
e^{\alpha}&=\chi^{-1}=\frac{a_6 \big(2 a_4+3 r^2\big)\big(a_4+2
r^2\big)}{\big(a_4+r^2\big) \big\{\big(2 a_4+3 r^2\big)
\big(a_6-r^2\big)+\zeta  c_2 a_6 r^2 \big(a_4+2 r^2\big)\big\}}.
\end{align}

Finally, the complexity factor \eqref{g54} takes the form for the
solution analogous to Eq.\eqref{g56b} by
\begin{align}\nonumber
\tilde{Y}_{TF}&=Y_{TF}=\frac{1}{80 a_6 r^3}\bigg[16 a_6 r^5 \omega
^2+20 (2 \beta +1) \big(a_4+2 a_6\big) r-45 \sqrt{2} \beta
\sqrt{a_4} \\\nonumber &\times \big(a_4+2 a_6\big)\tan
^{-1}\bigg(\frac{\sqrt{2} r}{\sqrt{a_4}}\bigg)+\frac{1}{\big(a_4+2
r^2\big)^2}\big\{10 (7 \beta -4)\big(a_4+2 r^2\big)
a_4\\\label{g60d} &\times \big(a_4+2 a_6\big) r-20 (\beta -1) a_4^2
\big(a_4+2 a_6\big) r\big\}\bigg].
\end{align}
Figure \textbf{7} depicts its plots by varying the Rastall parameter
and charge, as this is not dependent on $\zeta$. This factor admits
an increasing trend w.r.t. $r$, however, decreases by increasing the
parameter $\beta$. We also notice that this factor monotonically
increases within the range of $0 \leq r \leq 8$ and then again
decreases in contrast with the one discussed in Figure \textbf{6}.
\textcolor{blue} {Further, more the charge is, less complex is the
interior and thus indicating the significance of the presence of an
electromagnetic field in the study of compact stars.}
\begin{figure}[h!]\center
\epsfig{file=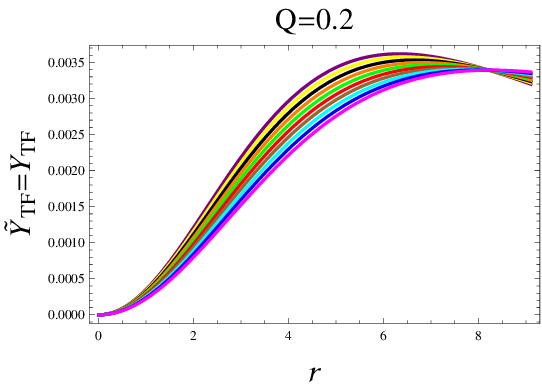,width=.38\linewidth}\epsfig{file=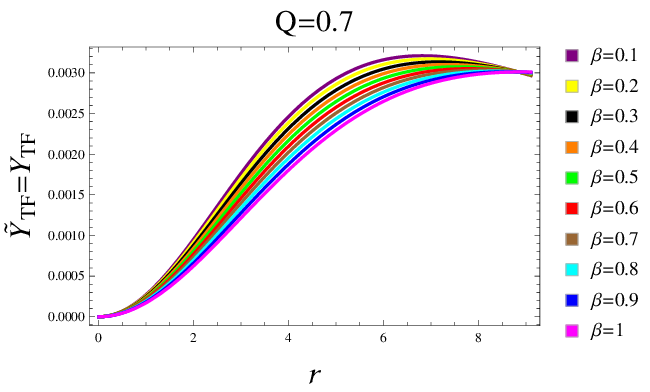,width=.45\linewidth}
\caption{Complexity factor \eqref{g60d}.}
\end{figure}

\subsection{Total Fluid Setup admitting Null Complexity}

Another constraint we choose in this subsection also depends on the
new gravitational source. It is considered that both the initial and
extra sources may have complexity totally opposite to each other,
i.e., when we add them, the complexity of the total fluid
distribution may vanish. This is mathematically expressed as
$Y^{\Theta}_{TF}\neq0$. On the other hand, we can say that the
initial source must have the non-null complexity factor, justifying
the utilization of the gravitational decoupling strategy that
simplifies the complex equations governing the geometry. When we
merge this constraint with Eqs.\eqref{g54}, \eqref{g57} and
\eqref{g58}, it becomes
\begin{align}\nonumber
&40 a_6 r^2 \big(a_4+r^2\big) \big(a_4+2 r^2\big) {d_f}'(r)+40 a_6 r
\big(a_4+r^2\big)^2 \big(r {d_f}'(r)-2 {d_f}(r)\big)\\\nonumber &-80
a_6 r \big(2 a_4 r^2+a_4^2+2 r^4\big)
{d_f}(r)+\big(a_4+r^2\big)^2\big\{16 a_6 r^5 \omega ^2+20 (2 \beta
+1)\\\nonumber &\times \big(a_4+2 a_6\big) r+\frac{10 (7 \beta -4)
a_4 \big(a_4+2 a_6\big) r}{a_4+2 r^2}-\frac{20 (\beta -1) a_4^2
\big(a_4+2 a_6\big) r}{\big(a_4+2 r^2\big)^2}\\\label{60e} &-45
\sqrt{2} \beta \sqrt{a_4} \big(a_4+2 a_6\big) \tan
^{-1}\big(\tau_3\big)\big\}=0,
\end{align}
where $\tau_3=\frac{\sqrt{2} r}{\sqrt{a_4}}$. Working out the
solution of the above equation for $d_f(r)$, we get the following
expression containing $c_3$ as a constant
\begin{align}\nonumber
d_f(r)&=\frac{c_3 r^2 \big(a_4+r^2\big)}{2 a_4+3
r^2}-\frac{a_4+r^2}{40 a_6 r \big(a_4+2 r^2\big) \big(2 a_4+3
r^2\big)}\big\{30 \beta r \big(a_4+2 a_6\big)
\\\nonumber &\times\big(a_4+2 r^2\big)-20 (\beta -1) \big(a_4+2 a_6\big) r^3
-8 a_6 r^5 \omega ^2 \big(a_4+2 r^2\big)\\\label{60f} &-15 \sqrt{2}
\beta \sqrt{a_4} \big(a_4+2 a_6\big) \big(a_4+2 r^2\big) \tan
^{-1}\big(\tau_3\big)\big\},
\end{align}
where $\mathrm{Dim}(c_3)=\frac{1}{\ell^2}$. We now use this with
Eq.\eqref{g17} to get modified radial component of the interior
geometry as
\begin{align}\nonumber
e^{\alpha}=\chi^{-1}&=\frac{40a_6r \big(a_4+2 r^2\big)\big(2a_4+3
r^2\big)}{a_4+r^2}\bigg[15 \sqrt{2} \beta  a_4^{5/2} \zeta  \tan
^{-1}\big(\tau _3\big)+4 a_4 r \big\{a_6\\\nonumber &\times \big(10
c_3 \zeta r^2-15 \beta \zeta+2 \zeta r^4 \omega ^2+20\big)-5 r^2 (2
\beta \zeta +\zeta +4)\big\}+8 a_6 r^3\\\nonumber &\times \big\{10
c_3 \zeta r^2+\zeta \big(2 r^4 \omega ^2-10 \beta
-5\big)+15\big\}-30 \beta  a_4^2 \zeta r-120 r^5+60 \\\label{g60fa}
&\times\sqrt{2} \beta \sqrt{a_4} a_6 \zeta  r^2 \tan ^{-1}\big(\tau
_3\big)+30 \sqrt{2} \beta a_4^{3/2} \zeta \big(a_6+r^2\big) \tan
^{-1}\big(\tau _3\big)\bigg]^{-1}.
\end{align}
This completes our minimally deformed solution analogous to the
constraint $\tilde{Y}_{TF}\neq0$. Finally, one can find the triplet
of fluid variables using Eqs.\eqref{g13} and \eqref{g60fa}. They are
presented in Appendix \textbf{A}.

The matter triplet \eqref{60g}-\eqref{60i} representing spherical
interior and its corresponding physical characteristics are now
interpreted through graphical representation. We again use the same
parametric variations as described earlier for the solution
corresponding to $\tilde{\Pi}=0$. Nonetheless, the constant $c_3$ is
differently adopted as $c_3=-0.045$. We firstly display the extended
$g_{rr}$ potential \eqref{g60fa} in Figure \textbf{8} whose minimum
value is 1 and maximum value varies when altering the Rastall and
decoupling parameters along with charge. Figures \textbf{9} and
\textbf{10} depict the matter triplet and anisotropy, and one can
easily observe variations in these factors w.r.t. the radial
coordinate and all other parameters. They are consistent with the
desired behavior, and thus with the first developed model as well.
\textcolor{blue} {The numerical values of the mass function for
certain parametric choices as $\beta=0.6$ and $\zeta=0.1$ are
$\tilde{M}=1.632M_{\bigodot}$ and $1.592M_{\bigodot}$ for $Q=0.2$
and $0.7$, respectively. However, the best fit mass with the
observed data is found for lower choices of the decoupling parameter
and higher values of the charge. The increasing anisotropy (depicted
in last rows of both Figures) also guarantees the physical
acceptability of this model.} The energy conditions which are
necessary to be checked in such scenarios are explored without
adding their plots, indicating the existence of usual fluid and
viability of the solution \eqref{60g}-\eqref{60i}. We eventually
check the stability in Figures \textbf{11} and \textbf{12}, and
reach at the conclusion that all plots are consistent with the
required criteria, leading our model to be stable.
\begin{figure}[h!]\center
\epsfig{file=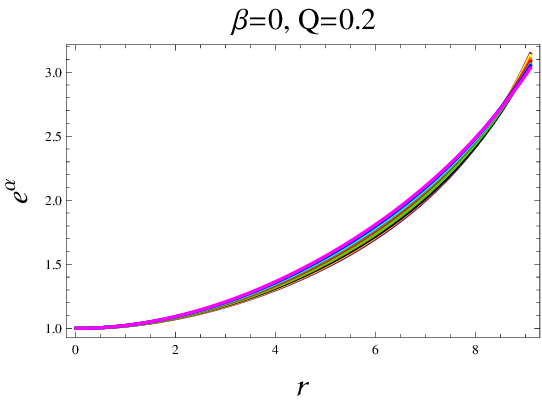,width=.35\linewidth}\epsfig{file=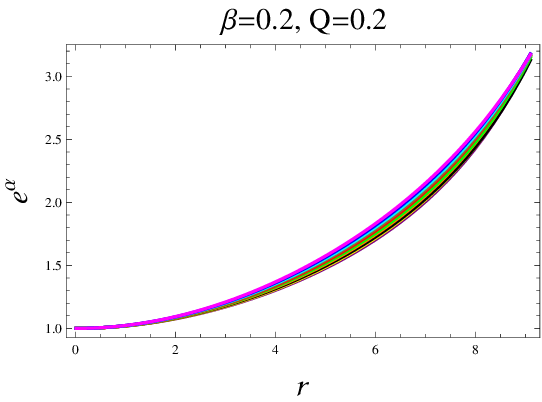,width=.35\linewidth}\epsfig{file=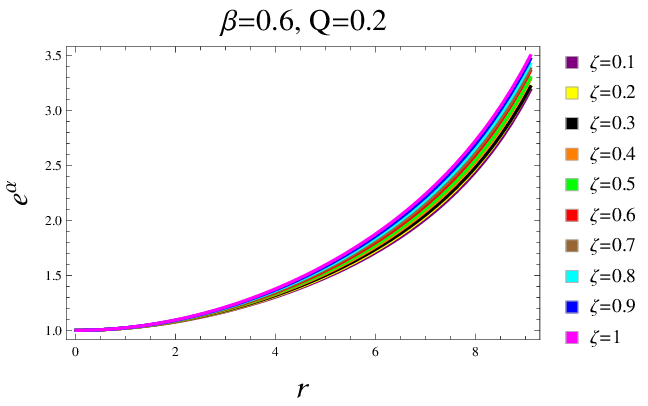,width=.42\linewidth}
\epsfig{file=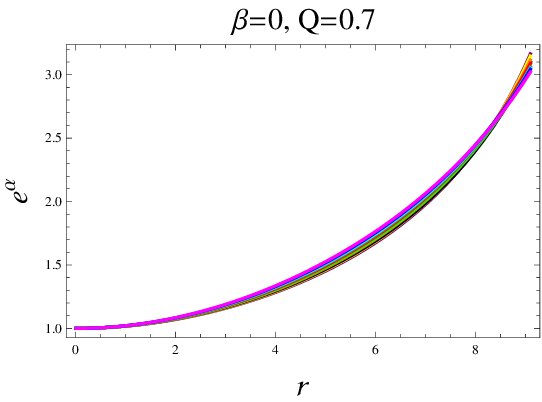,width=.35\linewidth}\epsfig{file=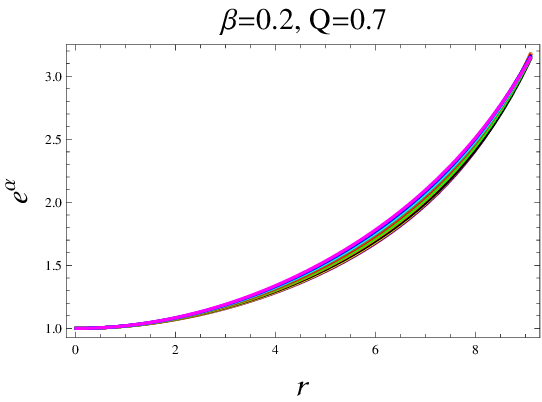,width=.35\linewidth}\epsfig{file=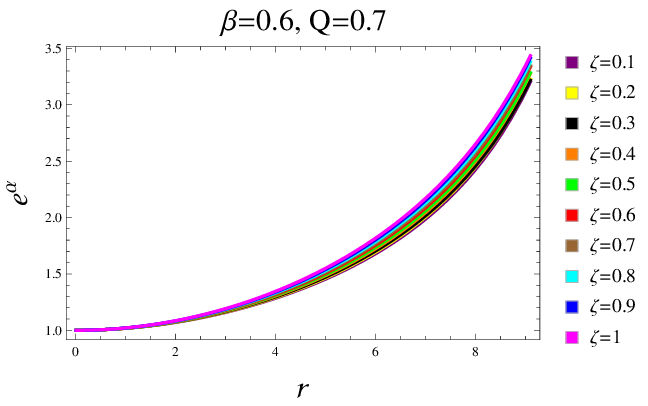,width=.42\linewidth}
\caption{Radial metric \eqref{g60fa}.}
\end{figure}
\begin{figure}[h!]\center
\epsfig{file=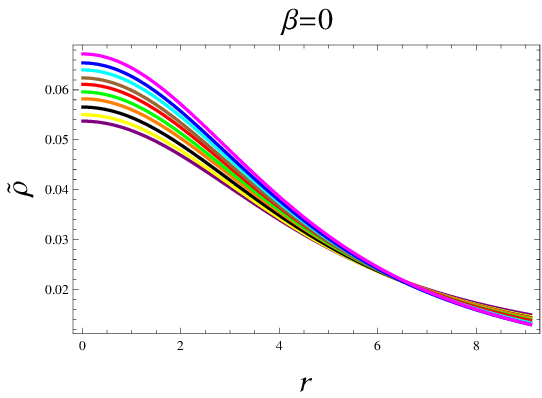,width=.35\linewidth}\epsfig{file=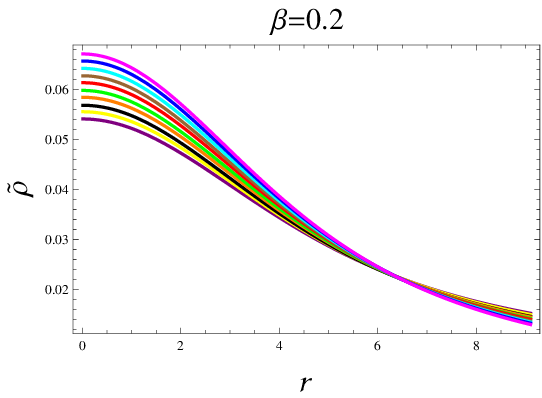,width=.35\linewidth}\epsfig{file=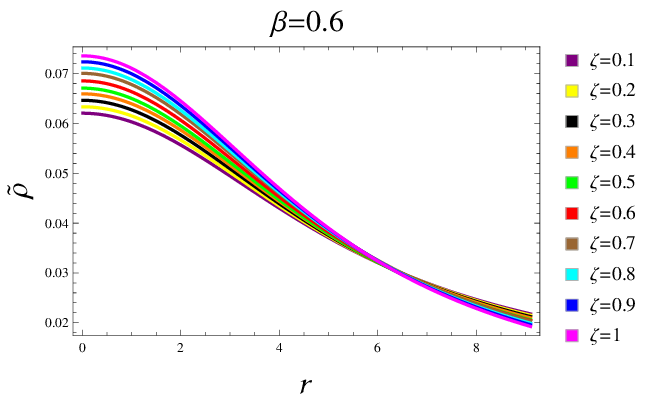,width=.42\linewidth}
\epsfig{file=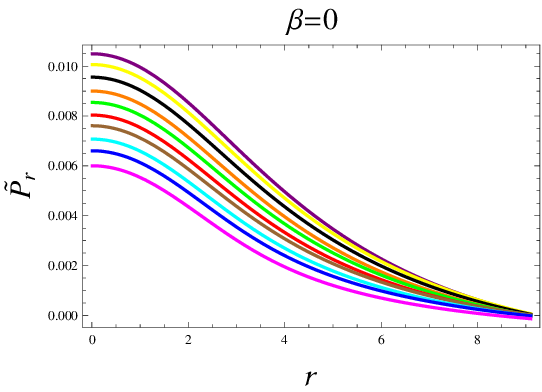,width=.35\linewidth}\epsfig{file=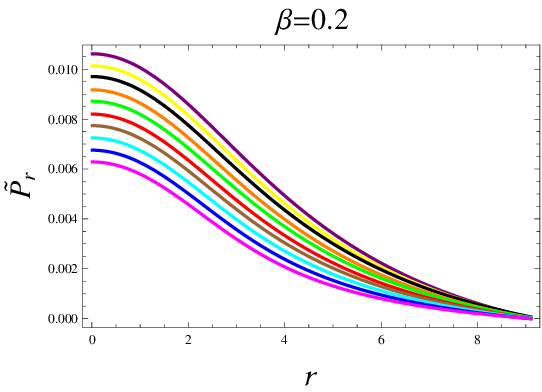,width=.35\linewidth}\epsfig{file=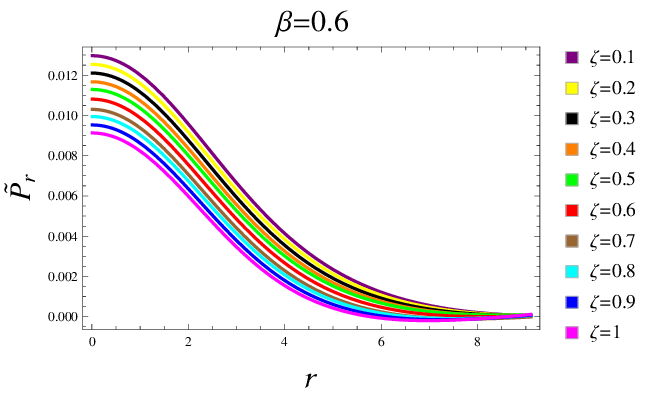,width=.42\linewidth}
\epsfig{file=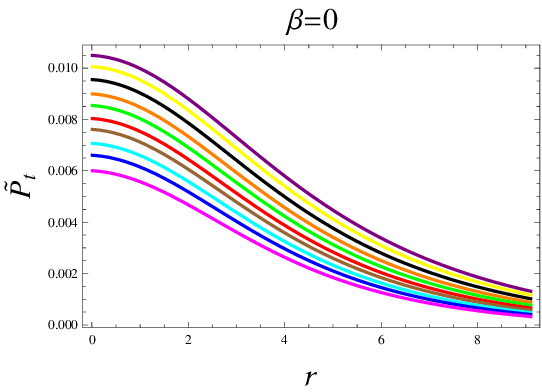,width=.35\linewidth}\epsfig{file=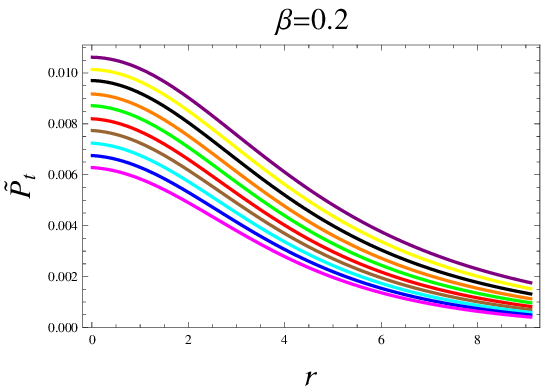,width=.35\linewidth}\epsfig{file=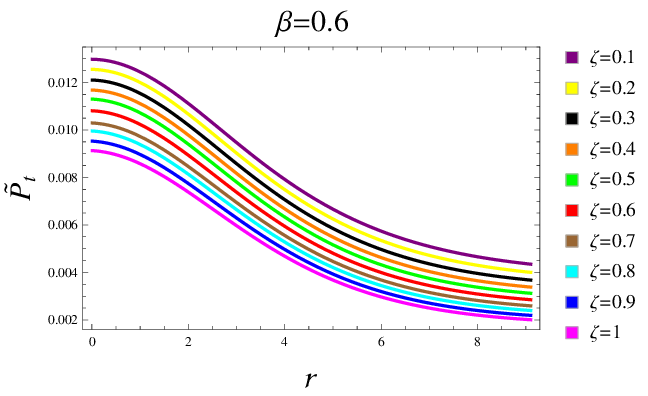,width=.42\linewidth}
\epsfig{file=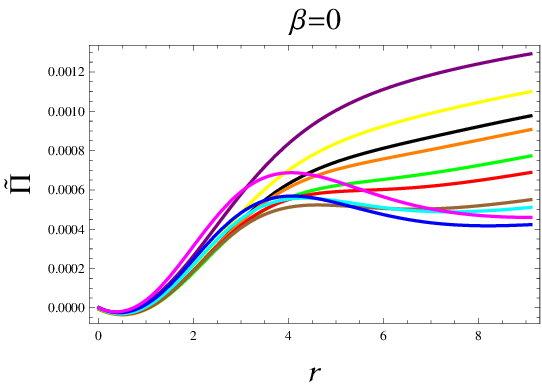,width=.35\linewidth}\epsfig{file=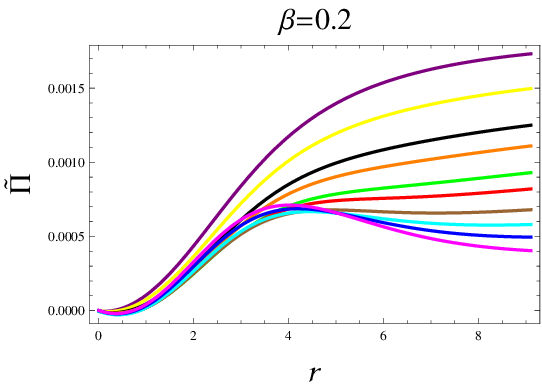,width=.35\linewidth}\epsfig{file=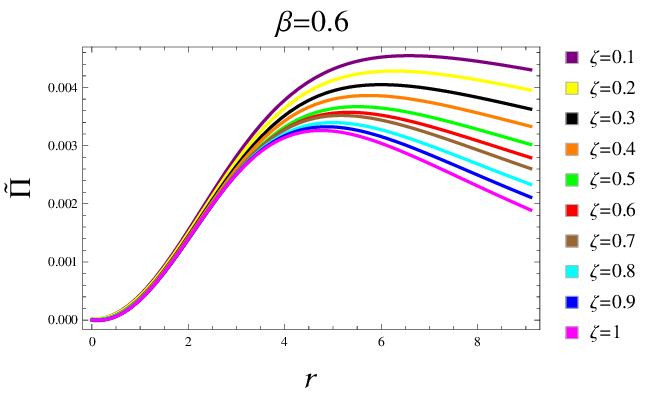,width=.42\linewidth}
\caption{Matter variables corresponding to $\mathrm{Q}=0.2$.}
\end{figure}
\begin{figure}[h!]\center
\epsfig{file=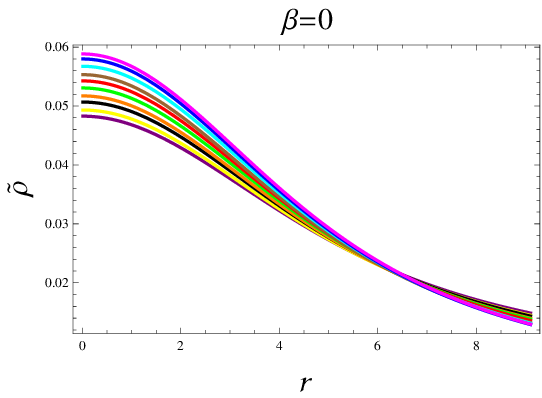,width=.35\linewidth}\epsfig{file=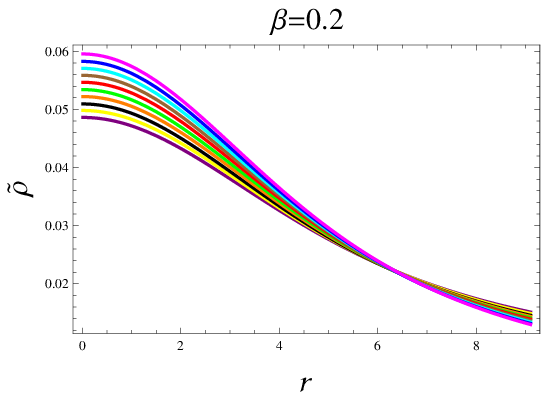,width=.35\linewidth}\epsfig{file=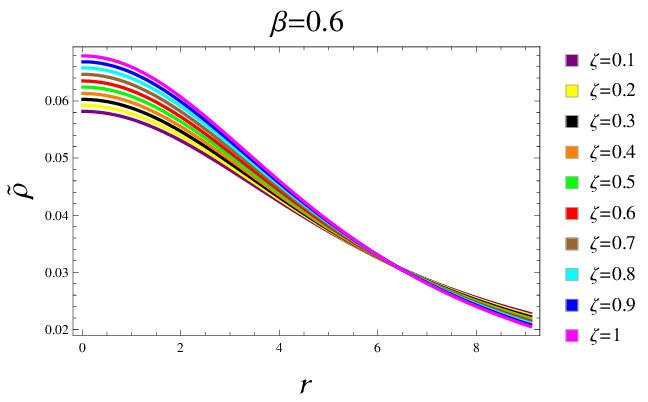,width=.42\linewidth}
\epsfig{file=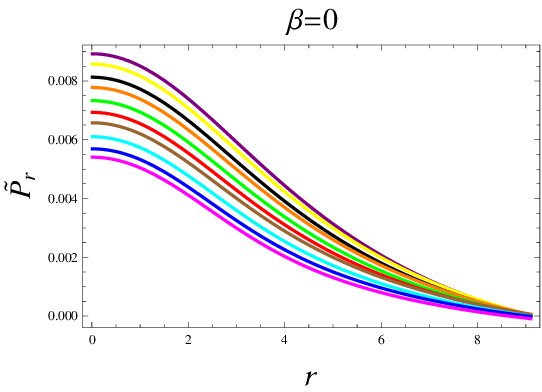,width=.35\linewidth}\epsfig{file=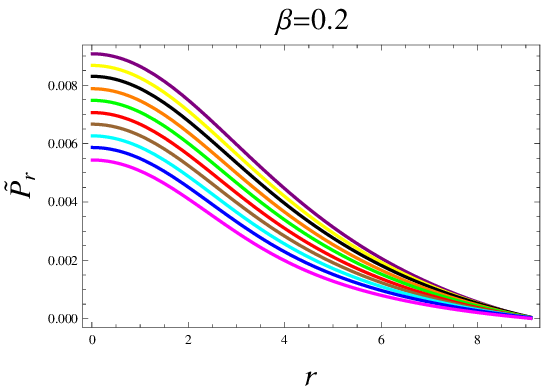,width=.35\linewidth}\epsfig{file=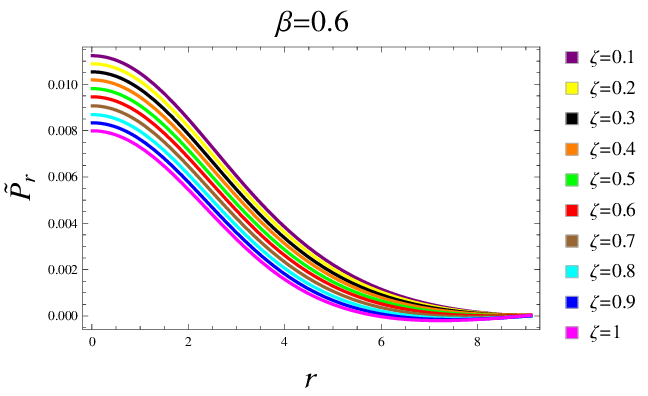,width=.42\linewidth}
\epsfig{file=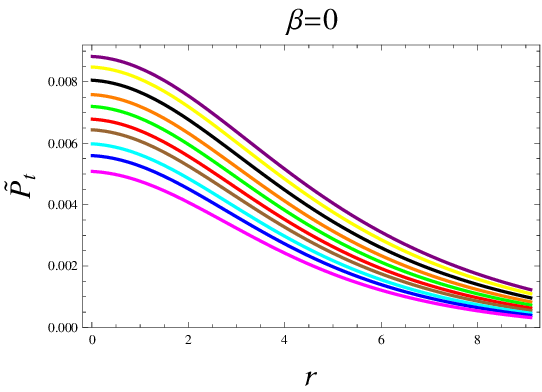,width=.35\linewidth}\epsfig{file=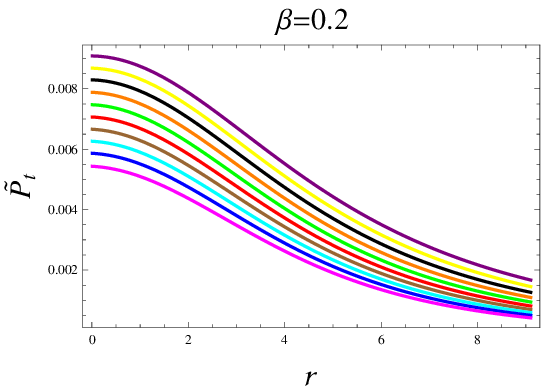,width=.35\linewidth}\epsfig{file=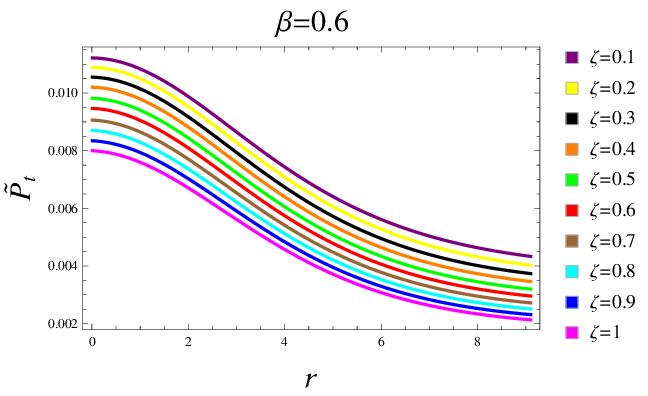,width=.42\linewidth}
\epsfig{file=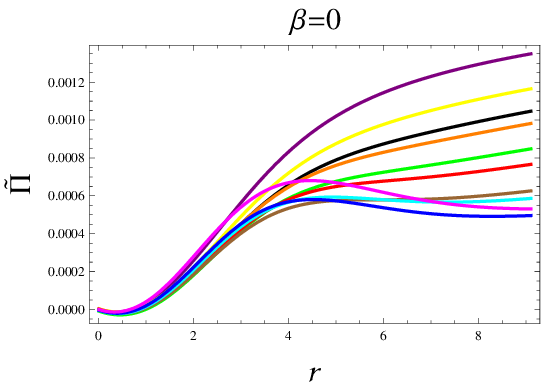,width=.35\linewidth}\epsfig{file=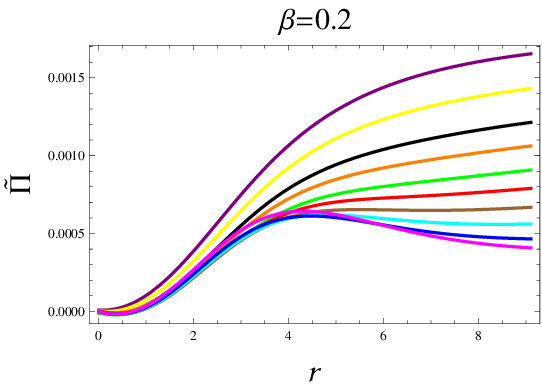,width=.35\linewidth}\epsfig{file=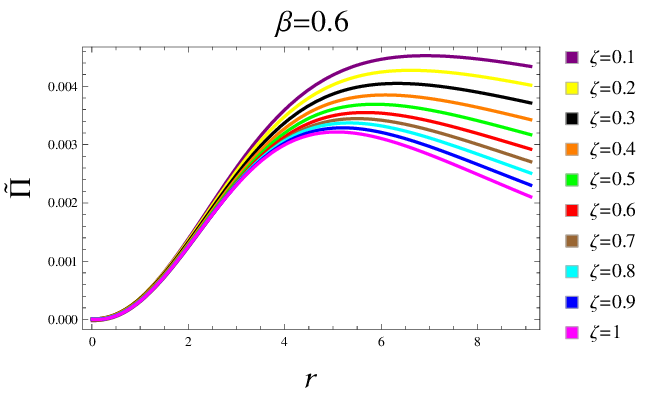,width=.42\linewidth}
\caption{Matter variables corresponding to $\mathrm{Q}=0.7$.}
\end{figure}
\begin{figure}[h!]\center
\epsfig{file=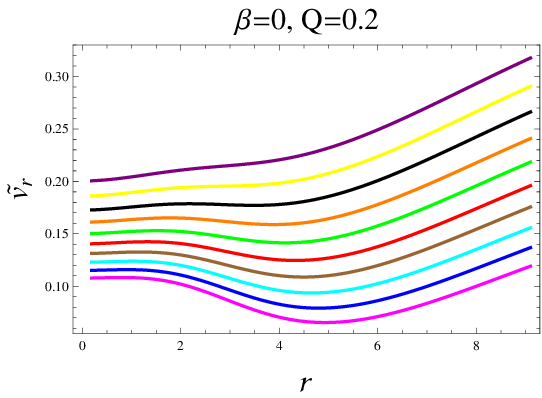,width=.35\linewidth}\epsfig{file=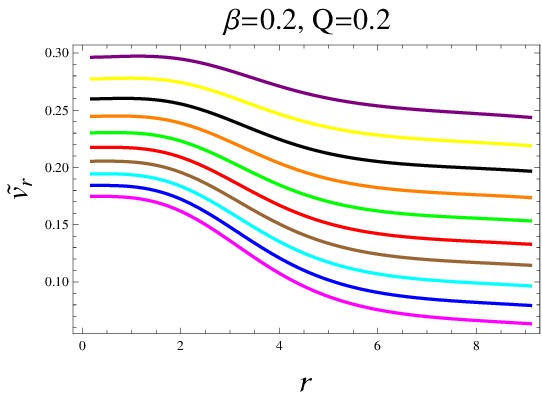,width=.35\linewidth}\epsfig{file=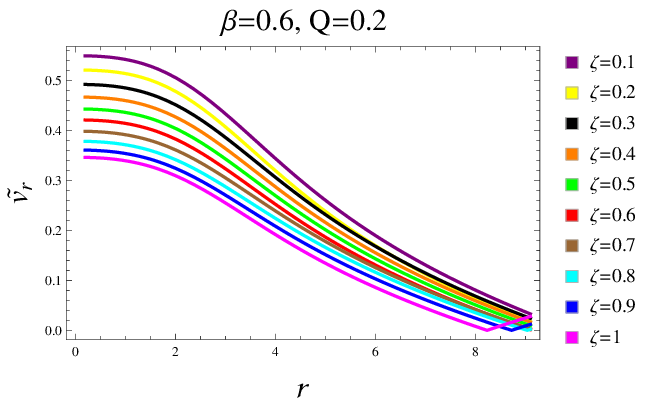,width=.42\linewidth}
\epsfig{file=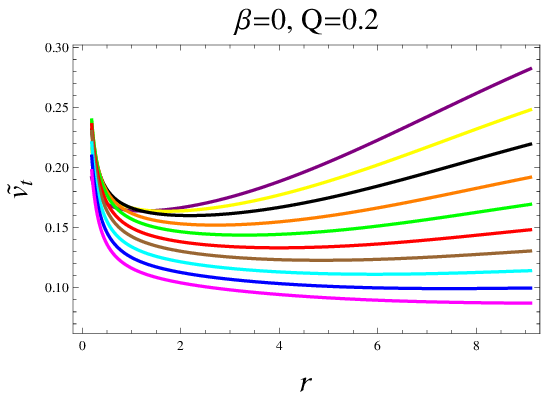,width=.35\linewidth}\epsfig{file=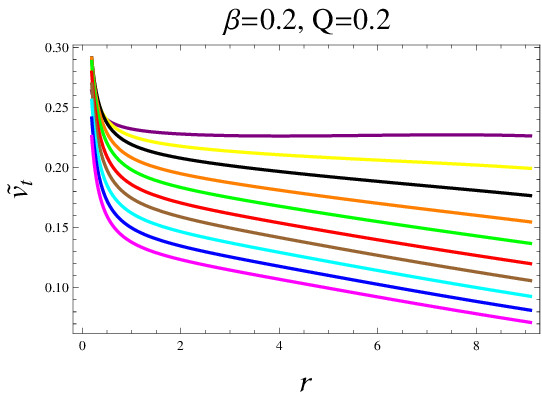,width=.35\linewidth}\epsfig{file=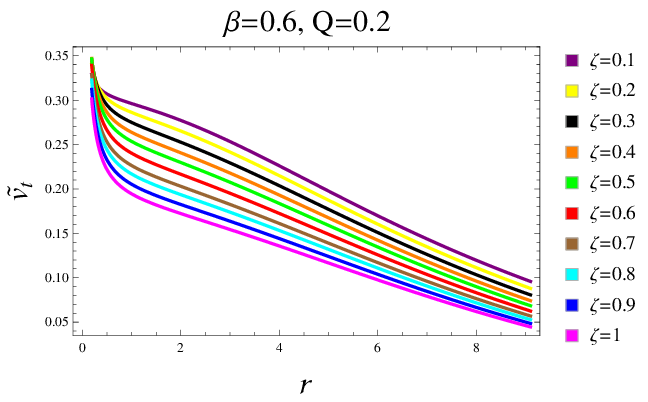,width=.42\linewidth}
\epsfig{file=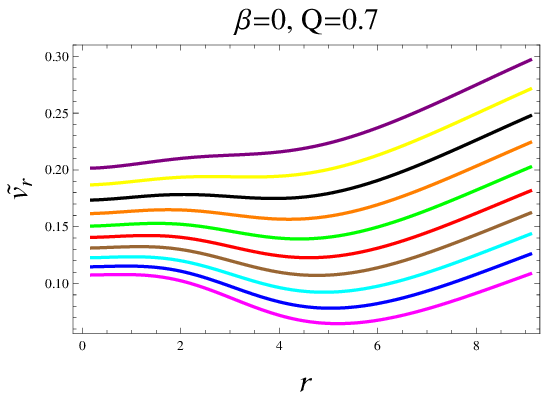,width=.35\linewidth}\epsfig{file=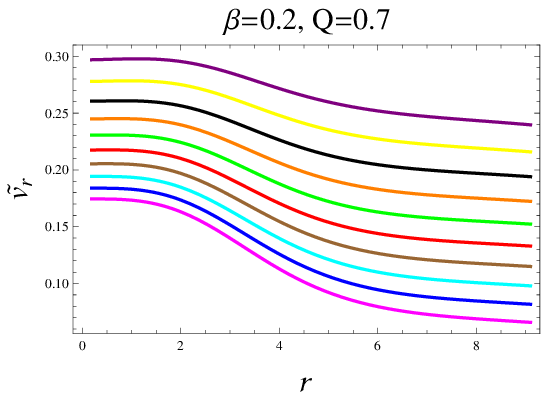,width=.35\linewidth}\epsfig{file=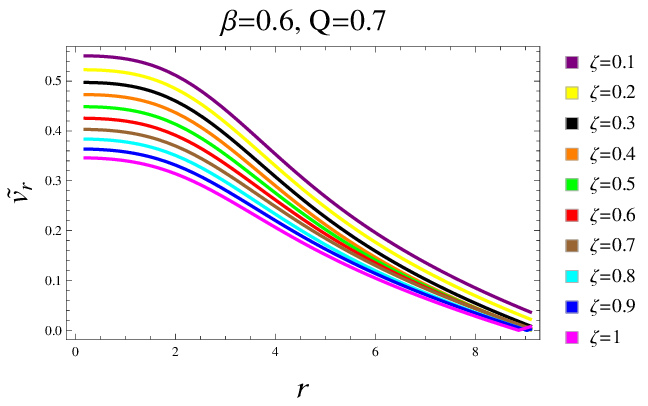,width=.42\linewidth}
\epsfig{file=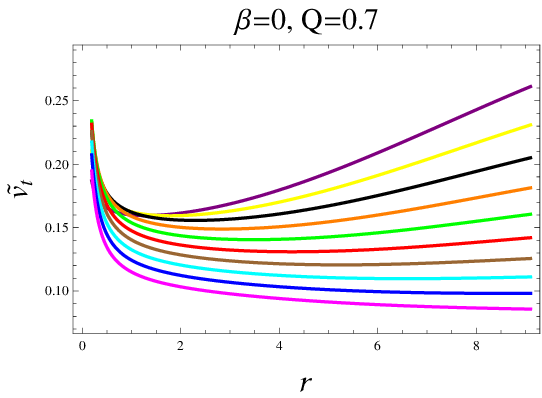,width=.35\linewidth}\epsfig{file=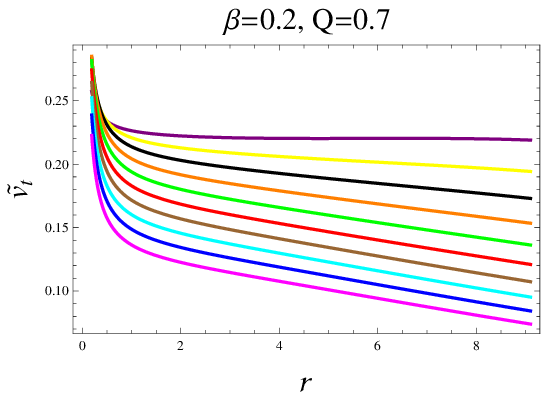,width=.35\linewidth}\epsfig{file=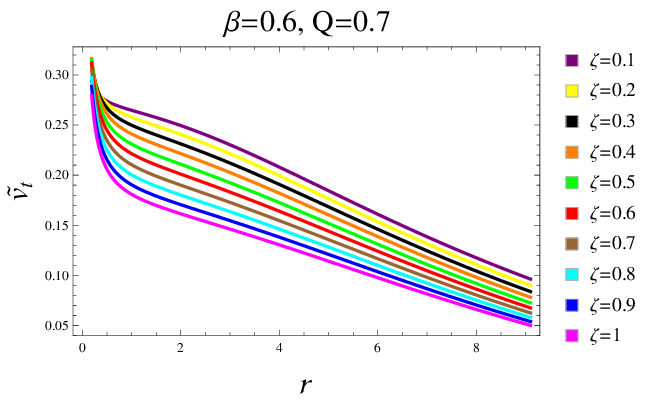,width=.42\linewidth}
\caption{Sound speeds in radial and tangential directions.}
\end{figure}
\begin{figure}[h!]\center
\epsfig{file=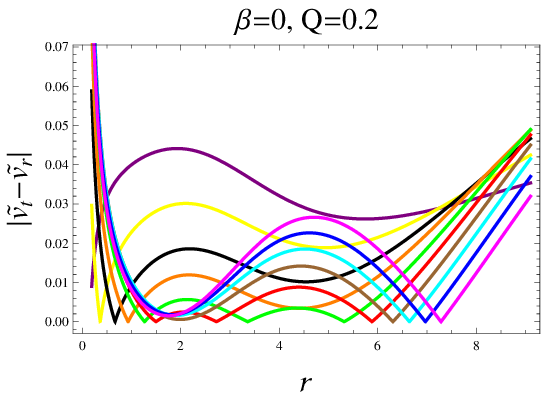,width=.35\linewidth}\epsfig{file=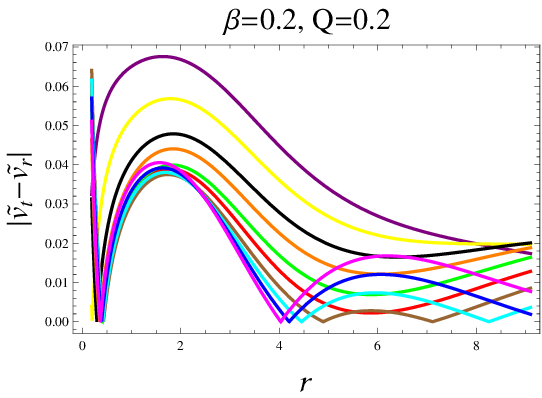,width=.35\linewidth}\epsfig{file=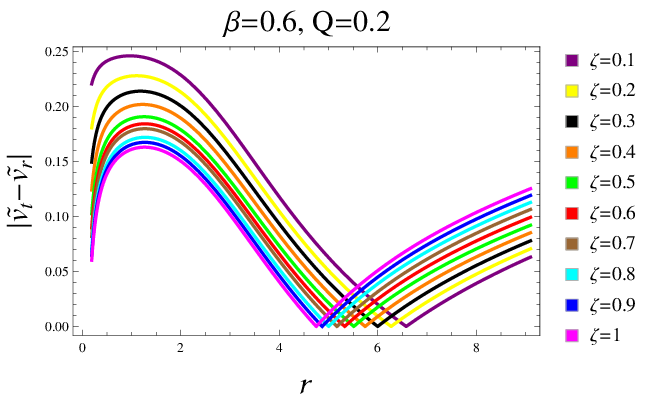,width=.42\linewidth}
\epsfig{file=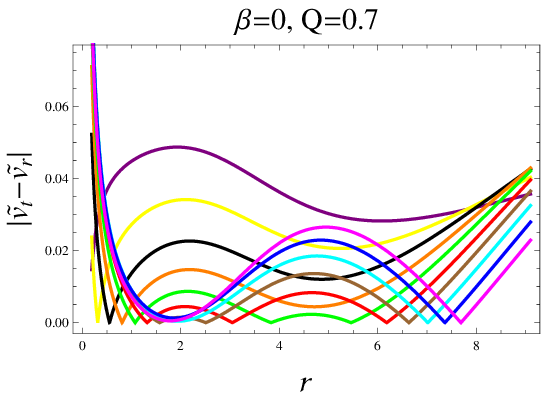,width=.35\linewidth}\epsfig{file=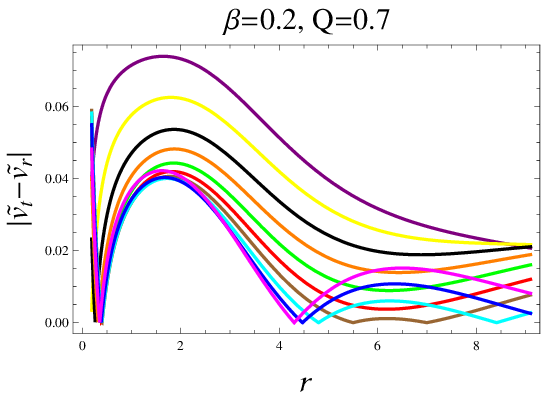,width=.35\linewidth}\epsfig{file=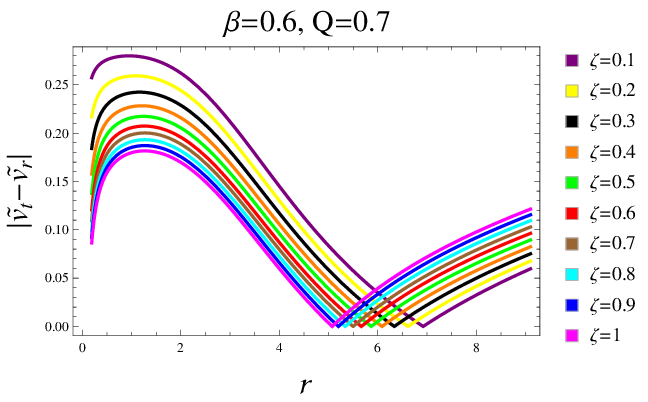,width=.42\linewidth}
\caption{Herrera's cracking criteria.}
\end{figure}

\section{Conclusions}

In this article, I have examined the established methodologies
within GR and applied them in the context of Rastall gravity theory.
This investigation commenced by adopting a charged static
anisotropic spherical interior and modifying it through the
incorporation of an additional source. This inclusion resulted in a
heightened intricacy within the Rastall gravitational equations, and
thus, to streamline and simplify them, I have employed the MGD
scheme. This method involved segregating the equations into separate
sets that describe specific fluid configurations. To work out the
field equations for an initial anisotropic fluid arrangement, two
distinct ansatz have been employed. These assumptions pertains to
the Krori-Barua spacetime configuration, which is defined by
$$\sigma(r)=a_1 r^2+a_2,\quad
\chi(r)=e^{-\alpha(r)}=e^{-a_3r^2},$$ while the second model being
Tolman IV metric potentials are provided as follows
$$\sigma(r)=\ln\bigg\{a_5\bigg(1+\frac{r^2}{a_4}\bigg)\bigg\},\quad
\chi(r)=e^{-\alpha(r)}=\frac{\big(a_4+r^2\big)\big(a_6-r^2\big)}{a_6\big(a_4+2r^2\big)}.$$
The triplet in each spacetime ansatz has been calculated using the
matching conditions at the interface. To achieve this, I utilized
observed radius and mass of a compact object, specifically $4U
1820-30$. It has also been observed that certain constants possess
distinct dimensions, while others are dimensionless. Furthermore,
the system represented by Eqs.\eqref{g21}-\eqref{g23} contained four
unknowns, necessitating the imposition of a single constraint on the
matter source $\Theta_{\eta\mu}$ at a time to get a solution. Three
distinct models have been formulated, each corresponding to a
specific constraint applied to the system as follows
\begin{itemize}
\item A mathematical isotropization of the anisotropic compact matter setup, i.e., $\tilde{\Pi}=0$ only
for a specific value of $\zeta$,

\item Vanishing complexity factor corresponding to extra matter setup, i.e.,
$Y_{TF}^{\Theta}=0$,

\item Vanishing complexity factor corresponding to total matter setup, i.e., $\tilde{Y}_{TF}=0$.
\end{itemize}

Subsequent to this, I have graphically analyzed the two models
corresponding to first and third constraints given above using
different Rastall, decoupling and charge parametric values. For
instance, I have adopted $\beta=0.2,0.6$, $\zeta=0.1,~0.2,~...~,1$
and $\mathrm{Q}=0.2,0.7$ to plot essential properties for the
formulated solutions. The corresponding modified radial metric
potential, effective form of fluid variables along with their
anisotropic factors have been assessed and found consistent with the
required behavior. It has further been observed that the decrement
in Rastall parameter and charge makes the interior more massive in
all cases. Afterwards, I have successfully displayed some graphical
plots to find compatibility of developed models with the viability
criteria \eqref{g50}, hence, they are composed of usual fluid.
\textcolor{blue} {Also, the interior geometry for the second model
possesses more mass as compared to the first one for all choices of
$\beta$ and $\zeta$.} Two complexity factors have been obtained in
Eqs.\eqref{g56a} and \eqref{g60d}, and plotted, showing increasing
behavior outwards, as can be confirmed from Figures \textbf{5} and
\textbf{6}. Finally, for the stability check, I have used two
techniques depending on the sound speed. \textcolor{blue} {It has
been confirmed that all these criteria produced stable and
physically existing models for all selected values of $\zeta$,
$\beta$ and charge. They are found to be consistent with their
analogues in GR \cite{37k}. It is also important to discuss here
that I find more interesting results in Rastall framework when
comparing them with the Brans-Dicke \cite{37m} and minimally coupled
$f(R,T)$ theory \cite{42d}. This is because the first solution is
stable for all parametric values only in the current setup. Extend
this work to include non-linear equations of state and examining the
effects of dark energy in future could further enrich our
understanding of cosmic evolution under non-conserved gravity
theories. Investigating the interplay between Rastall gravity and
other alternative theories may also unveil new avenues for
addressing current cosmological challenges, such as the $H_0$
tension and dark matter interactions.} Finally, these results shall
be reduced in GR after substituting $\beta=0$.

\section*{Appendix A}
\renewcommand{\theequation}{A\arabic{equation}}
\setcounter{equation}{0} The fluid parameters analogous to the
constraint $\tilde{Y}_{TF}\neq0$ are
\begin{align}\nonumber
\tilde{\rho}&=\frac{1}{a_6 \big(a_4+2 r^2\big)^2}\big\{2 a_6 r^2
\big(2 \beta -2 r^4 \omega ^2+1\big)-a_4^2 \big(a_6 r^2 \omega ^2+6
\beta -3\big)+a_4\big(a_6 \\\nonumber &\times  \big(3-4 r^4 \omega
^2\big)+(7-22 \beta ) r^2\big)+6 (1-4 \beta )
r^4\big\}-\frac{\zeta(2 a_4 r+3 r^3)^{-2} }{40 a_6 (\tau
_3^2+1)\big(a_4+2 r^2\big)^2}\\\nonumber &\times\big[4 a_4^2 r^4
\big(\tau _3^2+1\big) \big\{a_6 \big(850 c_3 r^2-175 \beta+294 r^4
\omega ^2-170\big)-15 (7 \beta +5) r^2\big\}\\\nonumber &+2 a_4^3
r^2 \big(\tau _3^2+1\big) \big\{a_6 \big(740 c_3 r^2-210 \beta +252
r^4 \omega ^2-120\big)-5 (35 \beta +34) r^2\big\}\\\nonumber &+10
a_4^4 \big(\tau _3^2+1\big) \big\{4 a_6 \big(6 c_3 r^2-3 \beta+2 r^4
\omega ^2\big)-3 (7 \beta +4) r^2\big\}+240 \big(\tau
_3^2+1\big)\\\nonumber &\times a_6 r^8 \big(6 c_3 r^2-2 \beta+2 r^4
\omega ^2-1\big)+8 a_4 r^6 \big(\tau _3^2+1\big) \big\{a_6 \big(440
c_3 r^2-105 \beta + r^4 \\\nonumber &\times 152\omega ^2-75\big)-15
(2 \beta +1) r^2\big\}+360 \sqrt{2} \beta \sqrt{a_4} a_6 r^8 \tau
_3'+15 \sqrt{2} \beta a_4^{7/2} r \big\{\big(26 a_6\\\nonumber &+31
r^2\big) r\tau _3'-4 \big(a_6+2 r^2\big) \big(\tau _3^2+1\big) \tan
^{-1}\big(\tau _3\big)\big\}+15 \sqrt{2} \beta a_4^{9/2}
\big\{\big(4 a_6+13 r^2\big) \\\nonumber &\times \tau _3'-2 r
\big(\tau _3^2+1\big) \tan ^{-1}\big(\tau _3\big)\big\}+60 \sqrt{2}
\beta a_4^{3/2} r^5 \big\{r \big(16 a_6+3 r^2\big) \tau
_3'-\big(\tau _3^2+1\big)\\\nonumber &\times 4 a_6 \tan
^{-1}\big(\tau _3\big)\big\}+30 \sqrt{2} \beta a_4^{5/2} r^3 \big\{r
\big(31 a_6+16 r^2\big) \tau _3'-4 \big(2 a_6+r^2\big) \big(\tau
_3^2+1\big)\\\label{60g} &\times \tan ^{-1}\big(\tau
_3\big)\big\}+30 \sqrt{2} \beta a_4^{11/2} \tau _3'-60 \beta a_4^5
\big(\tau _3^2+1\big)\big],\\\nonumber \tilde{P}_{r}&=\frac{1}{a_6
\big(a_4+2 r^2\big)^2}\big[a_4^2 \big(a_6 r^2 \omega ^2+6 \beta
-1\big)+2 r^2 \big\{a_6 \big(1-2 \beta +2 r^4 \omega
^2\big)+3r^2\\\nonumber &\times (4 \beta -1) \big\}+a_4 \big\{a_6
\big(4 r^4 \omega ^2+1\big)+(22 \beta -5)
r^2\big\}\big]+\frac{\zeta\big(3r^2+a_4\big)}{r^2+a_4}\bigg[c_3
r^2\\\nonumber &\times\frac{ \big(a_4+r^2\big)}{2 a_4+3
r^2}-\frac{a_4+r^2}{40 a_6 r \big(a_4+2 r^2\big) \big(2 a_4+3
r^2\big)}\big\{30 \beta r \big(a_4+2 a_6\big)\big(a_4+2 r^2\big)
\\\nonumber &-20 (\beta -1) \big(a_4+2 a_6\big) r^3
-8 a_6 r^5 \omega ^2 \big(a_4+2 r^2\big)-15 \big(a_4+2 a_6\big)
\big(a_4+2 r^2\big)\\\label{60h} & \times \sqrt{2} \beta
\sqrt{a_4}\tan ^{-1}\big(\tau_3\big)\big\}\bigg],\\\nonumber
\tilde{P}_{t}&=\frac{1}{a_6 \big(a_4+2 r^2\big)^2}\big[a_4^2
\big(a_6 r^2 \omega ^2+6 \beta -1\big)+2 r^2 \big\{a_6 \big(1-2
\beta +2 r^4 \omega ^2\big)+3r^2\\\nonumber &\times (4 \beta -1)
\big\}+a_4 \big\{a_6 \big(4 r^4 \omega ^2+1\big)+(22 \beta -5)
r^2\big\}\big]+\frac{\zeta}{80 a_6 r^3 (2 a_4+3 r^2)^2}\\\nonumber
&\times\frac{1}{(a_4+r^2)(a_4+2 r^2)^{2}}\big[4 a_4^4 r^3 \big\{2
a_6 \big(20 c_3 (9 r+4) r-25 \beta+66 r^4 \omega ^2+32 r^3 \omega
^2\\\nonumber &-20\big)-5 r ((37 \beta +20) r-28 \beta+16)\big\}+4
a_4^5 r^3 \big\{8 a_6 \big(5 c_3+2 r^2 \omega ^2\big)-(5 \beta
+4)\\\nonumber &\times 5\big\}+96 a_6 r^{11} \big\{-10 \beta +20 c_3
(r+2) r+6 r^4 \omega ^2+16 r^3 \omega ^2-5\big\}+4 a_4^3 r^4 \big\{2
a_6 \\\nonumber &\times \big(20 (7 \beta -4)+30 c_3 (21 r+16)
r^2+214 r^5 \omega ^2+200 r^4 \omega ^2-5 (37 \beta +20)
r\big)\\\nonumber &-5 r^2 ((87 \beta +42) r-80 \beta+32)\big\}+8
a_4^2 r^6 \big\{a_6 \big(80 (5 \beta -2)+20 c_3 (53 r+54)\\\nonumber
& \times r^2+338 r^5 \omega ^2+464 r^4 \omega ^2-15 (29 \beta +14)
r\big)-20 r^2 (5 (2 \beta +1) r-7 \beta+1)\big\}\\\nonumber &+16 a_4
r^8 \big\{4 a_6 \big(35 \beta +35 c_3 (3 r+4) r^2+32 r^5 \omega
^2+60 r^4 \omega ^2-25 (2 \beta +1) r-5\big)\\\nonumber &-15 (2
\beta +1) r^3\big\}+360 \sqrt{2} \beta \sqrt{a_4} a_6 r^9 \big(r
(r+4) \tau _4'+(r-4) \tau _4\big)+30 \sqrt{2} \beta a_4^{9/2} r
\\\nonumber &\times \big\{r \big(a_6 (15 r+8)+2 (11 r+13) r^2\big) \tau
_4'-\big(a_6 (9 r+8)+3 (r+10) r^2\big) \tau _4\big\}+60\\\nonumber
&\times \sqrt{2} \beta a_4^{3/2} r^7 \big\{r \big(a_6 (22 r+64)+3
(r+4) r^2\big) \tau _4'+\big(2 a_6 (9 r-40)+3 (r-4)
r^2\big)\\\nonumber &\times \tau _4\big\}+30 \sqrt{2} \beta
a_4^{5/2} r^5 \big\{r \big(a_6 (63 r+124)+2 (11 r+32) r^2\big) \tau
_4'+\big( (23 r-156)\\\nonumber &\times a_6+2 (9 r-40) r^2\big) \tau
_4\big\}+15 \sqrt{2} \beta a_4^{7/2} r^3 \big\{r \big(8 a_6 (11
r+13)+(63 r+124) r^2\big) \\\nonumber &\times \tau _4'+\big(r^2 (23
r-156)-12 a_6 (r+10)\big) \tau _4\big\}-30 \sqrt{2} \beta a_4^{13/2}
\big(\tau _4-r \tau _4'\big)+15\beta\\\label{60i} & \times
\sqrt{2}a_4^{11/2} \big\{r \big(4 a_6+r (15 r+8)\big) \tau
_4'-\big(4 a_6+r (9 r+8)\big) \tau _4\big\}\big],
\end{align}
where $\tau _4=\tan ^{-1}(\tau _3)$. Moreover, the last two
equations expressed above are used to find the corresponding
anisotropy as
\begin{align}\nonumber
\tilde{\Pi}&=\frac{\zeta}{80 a_6 r^3 (a_4+r^2)(a_4+2 r^2)^{2}(2
a_4+3 r^2)^2}\big[4 a_4^4 r^3 \big\{2 a_6 \big(20 c_3 (9 r+4) r-25
\beta\\\nonumber &+66 r^4 \omega ^2+32 r^3 \omega ^2-20\big)-5 r
((37 \beta +20) r-28 \beta+16)\big\}+4 a_4^5 r^3 \big\{8 a_6 \big(5
c_3\\\nonumber &+2 r^2 \omega ^2\big)-(5 \beta +4)5\big\}+96 a_6
r^{11} \big\{-10 \beta +20 c_3 (r+2) r+6 r^4 \omega ^2+16 r^3 \omega
^2\\\nonumber &-5\big\}+4 a_4^3 r^4 \big\{2 a_6\big(20 (7 \beta
-4)+30 c_3 (21 r+16) r^2+214 r^5 \omega ^2+200 r^4 \omega
^2-5\\\nonumber &\times (37 \beta +20) r\big)-5 r^2 ((87 \beta +42)
r-80 \beta+32)\big\}+8 a_4^2 r^6 \big\{a_6 \big(20 c_3 (53
r+54)\\\nonumber &+80 (5 \beta -2)r^2+338 r^5 \omega ^2+464 r^4
\omega ^2-15 (29 \beta +14) r\big)-20 r^2 (5 (2 \beta +1)
r\\\nonumber & -7 \beta+1)\big\}+16 a_4 r^8 \big\{4 a_6 \big(35
\beta +35 c_3 (3 r+4) r^2+32 r^5 \omega ^2+60 r^4 \omega ^2-25
r\\\nonumber &\times (2 \beta +1) -5\big)-15 (2 \beta +1)
r^3\big\}+360 \sqrt{2} \beta \sqrt{a_4} a_6 r^9 \big(r (r+4) \tau
_4'+(r-4) \tau _4\big)\\\nonumber &+30 \sqrt{2} \beta a_4^{9/2}
r\big\{r \big(a_6 (15 r+8)+2 (11 r+13) r^2\big) \tau _4'-\big(a_6 (9
r+8)+3 (r+10)\\\nonumber &\times r^2\big) \tau _4\big\}+60\sqrt{2}
\beta a_4^{3/2} r^7 \big\{r \big(a_6 (22 r+64)+3 (r+4) r^2\big) \tau
_4'+\big(2 a_6 (9 r-40)\\\nonumber &+3 (r-4) r^2\big)\tau
_4\big\}+30 \sqrt{2} \beta a_4^{5/2} r^5 \big\{r \big(a_6 (63
r+124)+2 (11 r+32) r^2\big) \tau _4'+\big(a_6\\\nonumber & \times
(23 r-156)+2 (9 r-40) r^2\big) \tau _4\big\}+15 \sqrt{2} \beta
a_4^{7/2} r^3 \big\{r \big(8 a_6 (11 r+13)+(63 r\\\nonumber &+124)
r^2\big) \tau _4'+\big(r^2 (23 r-156)-12 a_6 (r+10)\big) \tau
_4\big\}-30 \sqrt{2} \beta a_4^{13/2} \big(\tau _4-r \tau
_4'\big)\\\nonumber &+15 \sqrt{2}a_4^{11/2}\beta \big\{r \big(4
a_6+r (15 r+8)\big) \tau _4'-\big(4 a_6+r (9 r+8)\big) \tau
_4\big\}\big]-\frac{\zeta\big(3r^2+a_4\big)}{r^2+a_4}\\\nonumber
&\times\bigg[\frac{c_3 r^2\big(a_4+r^2\big)}{2 a_4+3
r^2}-\frac{a_4+r^2}{40 a_6 r \big(a_4+2 r^2\big) \big(2 a_4+3
r^2\big)}\big\{30 \beta r \big(a_4+2 a_6\big)\big(a_4+2 r^2\big)
\\\nonumber &-20 (\beta -1) \big(a_4+2 a_6\big) r^3
-8 a_6 r^5 \omega ^2 \big(a_4+2 r^2\big)-15 \big(a_4+2 a_6\big)
\big(a_4+2 r^2\big)\\\label{60j} & \times \sqrt{2} \beta
\sqrt{a_4}\tan ^{-1}\big(\tau_3\big)\big\}\bigg].
\end{align}

\end{document}